\documentclass[aps,prd,preprint,groupedaddress,showpacs]{revtex4}

\usepackage{epsfig}
\usepackage{graphics}
\usepackage{slashed}
\usepackage{color}
\usepackage[citecolor=true]{hyperref}
\usepackage{multirow}
\usepackage{amsmath}
\begin{document}

\leftmargin -2cm
\def\choosen{\atopwithdelims..}

\boldmath
\title{Single and pair $J/\psi$ production in the Improved Color Evaporation Model \\
using the Parton Reggeization Approach} \unboldmath

\author{\firstname{A.A.}\surname{ Chernyshev}} \email{aachernyshoff@gmail.com}
\affiliation{Samara National Research University,  Samara, 443086,
Russia}

\author{\firstname{V.A.}\surname{ Saleev}} \email{saleev@samsu.ru}
\affiliation{Samara National Research University, Samara, 443086,
Russia} \affiliation{Joint Institute for Nuclear Research, Dubna,
141980 Russia}

\begin{abstract}
In the article, we study single and pair $J/\psi$ hadroproduction in
the Improved Color Evaporation Model via the Parton Reggeization
Approach. The last one is based on $k_T$--factorization of hard
processes in multi-Regge kinematics, the Kimber-Martin-Ryskin-Watt
model for unintegrated parton distribution functions, and the
effective field theory of Reggezied gluons and quarks, suggested by
L.N.~Lipatov. We compare contributions from the single and double
parton scattering mechanisms in the pair $J/\psi$ production. The
numerical calculations are realized using the Monte-Carlo event
generator KaTie. \vspace{0.2cm}
\end{abstract}

\maketitle

\section{Introduction}
Hadroproduction of $J/\psi$--mesons is being intensively studied
theoretically and experimentally for more than 50 years, after their
discovery in 1974. Experimental data on single $J/\psi$ production
are obtained in a wide range of the energy from $\sqrt{s} = 19$ GeV
up to $\sqrt{s} = 13$
TeV~\cite{NA3,AFS,PHENIX,Tevatron,LHCb1a,ALICE,ATLAS,CMS,LHCb1b}.
The processes of pair production of $J/\psi$ mesons were studied in
experiments at the Large Hadron Collider (LHC) by the
CMS~\cite{CMS_2Psi}, ATLAS~\cite{ATLAS_2Psi} and
LHCb~\cite{LHCb_2Psi} collaborations at energies of $7$, $8$, and
$13$ TeV. The theoretical description of the processes of charmonium
production is based on the perturbation theory of quantum
chromodynamics (QCD) in the constant of strong interaction
$\alpha_S(\mu)$, where the hard
 scale $\mu\sim m_\psi$, $m_\psi$ is the charmonium mass, and $\alpha_S \simeq 0.2$.
Hadronization process of the $c \bar{c}$-pair to the charmonium is
also described in terms of perturbation theory, only by the relative
velocity of $c(\bar{c})$ quarks in the charmonium. It is implemented
in the nonrelativistic quantum chromodynamics (NRQCD) approach
\cite{NRQCD}. In the leading approximation of the NRQCD, a quark and
an antiquark are produced in the color singlet state, as assumed in
the color singlet model (CSM)\cite{CSM1,CSM2}. Despite the success
of NRQCD in describing the charmonium spectra and cross-sections at
high energies, there are still unsolved problems: description of
$\eta_c-$meson production with allowance for the octet contribution
of NRQCD leads to an excess of predictions over experimental
data~\cite{etac-NLO}; NRQCD predicts that prompt $J/\psi$ should be
produced mostly transversely polarized, experimentally this is not
observed~\cite{NRQCD_spin}. The latter ones may indicate the
essential role of such nonperturbative effects that are not taken
into account in the NRQCD. Alternative, but the more
phenomenological approach is the Color Evaporation Model (CEM)
proposed many years ago in Ref. \cite{CEM1,CEM2}. Later, the CEM was
improved by Ma and Vogt~\cite{ICEM_Ma_Vogt} and now is used to
describe the spectra and polarizations of $J/\psi$-mesons in the
collinear parton model (CPM)~\cite{ICEM2017,ICEM2021} and in the
$k_T$-factorization approach~\cite{ICEM_KT1,ICEM_KT2}.

In the present study, we calculate the transverse momentum spectra
of prompt $J/\psi$ within framework of the  High Energy
Factorization (HEF) or the $k_T-$factorization, which initially has
been introduced as a resummation tool for $\ln
(\sqrt{s}/\mu)$-enhanced corrections to the hard-scattering
coefficients in the CPM, where invariant $\sqrt{s}$ referees to the
total energy of the process \cite{KT1,KT2,Gribov}. We use the parton
Reggeization approach (PRA) which is a version of HEF formalism,
based on the modified Multi-Regge Kinematics (MRK) approximation for
QCD scattering amplitudes~\cite{NSS2013,NKS2017,NefedovSaleev2020}.
The PRA is accurate both in the collinear limit, which drives the
Transverse-Momentum-Dependent (TMD) factorization \cite{CPM} and in
the high-energy (Multi-Regge) limit, which is important for
Balitsky-Fadin-Kuraev-Lipatov~(BFKL)~\cite{BFKL1,BFKL2,BFKL3,BFKL4}
resummation of $\ln (\sqrt{s}/\mu) $-enhanced effects. In the PRA,
we have studied successfully a heavy quarkonium production in the
proton-(anti)proton collisions at the Tevatron and the LHC using
NRQCD approach, see Refs.\cite{VKS2005A,VKS2005B,NSS2012,NS2016}.

The paper has the following structure. In Section~\ref{sec:PRA}, the
relevant basics of the PRA formalism are outlined. The Improved
Color Evaporation Model (ICEM) is shortly reviewed in Section
\ref{sec:icem}. In Section~\ref{sec:methods}, we overview
Monte-Carlo (MC) parton-level event generator KaTie~\cite{katie} and
the relation between calculations via the PRA and KaTie for
tree-level amplitudes. In Section~\ref{sec:results1}, we describe
the experimental data for the single prompt $J/\psi$ production at
the energy range from the $19$ GeV up to $13$ TeV.  In
Section~\ref{sec:results2}, we describe the experimental data for
the pair prompt $J/\psi$ production at the energy of the LHC
collider. Our conclusions are summarized in
Section~\ref{sec:conclusions}.

\section{Parton Reggeization Approach}
\label{sec:PRA} The PRA is based on the factorization hypothesis of
the HEF or $k_T$--factorization justified in the leading logarithmic
approximation of the QCD at high energies~\cite{KT1,KT2,Gribov}.
Dependent on transverse momentum, parton distribution functions
(PDF) of Reggeized quarks and gluons are calculated in the model
proposed earlier by Kimber, Martin, Ryskin and Watt
(KMRW)\cite{KMR,WMR}, but with sufficient
modifications~\cite{NefedovSaleev2020} that will be described below.
Reggeized parton amplitudes are constructed according to the Feynman
rules of the L.N. Lipatov Effective Field Theory (EFT) of Reggeized
gluons and quarks~\cite{Lipatov95, LipatovVyazovsky}. A detailed
description of the PRA can be found in
Refs.~\cite{NSS2013,NKS2017,NefedovSaleev2020}, inclusion of
corrections from the emission of additional partons to the leading
PRA approximation was studied in the Refs.~\cite{NKS2017, NS2019},
the development of the PRA with loop corrections was considered in
the Refs~\cite{PRANLO1,PRANLO2,PRANLO3}.

In the PRA, the cross-section of the process $p + p \to J/\psi + X$
is related to the cross-section of the parton subprocess by the
factorization formula
\begin{eqnarray}
  d\sigma & = & \sum_{i, \bar{j}}
    \int\limits_{0}^{1} \frac{dx_1}{x_1} \int \frac{d^2{\bf q}_{T1}}{\pi}
{\Phi}_i(x_1,t_1,\mu^2)
    \int\limits_{0}^{1} \frac{dx_2}{x_2} \int \frac{d^2{\bf q}_{T2}}{\pi}
{\Phi}_{j}(x_2,t_2,\mu^2)\cdot d\hat{\sigma}_{\rm PRA},
\label{eqI:kT_fact}
\end{eqnarray}
where $t_{1,2} = - {\bf{q}}_{T 1,2}^2$, the cross-section of the
subprocess with Reggeized partons $\hat{\sigma}_{\mathrm{PRA}}$ is
expressed in terms of squared Reggeized amplitudes
$\overline{|{\mathcal{A}}_{\mathrm{PRA}}|^2}$ in the standard way.

The PRA hard-scattering amplitudes are gauge-invariant because the
initial-state off-shell partons are considered as Reggeized partons
of the gauge-invariant EFT for QCD processes in the MRK limit
~\cite{Lipatov95,LipatovVyazovsky}. The Feynman rules of the Lipatov
EFT are written down in the
Refs.~\cite{LipatovVyazovsky,AntonovLipatov}. The easy way to use
Feynman rules of Lipatov EFT is the exploration a model file
ReggeQCD~\cite{NKS2017} for the FeynArts tool~\cite{hahn}.

Unintegrated PDFs (unPDFs) in the modified KMRW model are calculated
by the formula~\cite{NefedovSaleev2020}
\begin{equation}
\Phi_i(x,t,\mu) = \frac{\alpha_s(\mu)}{2\pi}
\frac{T_i(t,\mu^2,x)}{t}
\sum\limits_{j = q,\bar{q},g}\int\limits_{x}^{1} dz\ P_{ij}(z) {F}_j
\left(
\frac{x}{z}, t \right) \theta\left( \Delta(t,\mu)-z
\right),\label{uPDF}
\end{equation}
where $F_i(x,\mu_F^2) = x f_j(x,\mu_F^2)$. Here and below, we put
factorization and renormalization scales are equal, $\mu_F = \mu_R =
\mu$, and $\Delta(t,\mu^2)=\sqrt{t}/(\sqrt{\mu^2}+\sqrt{t})$ is the
KMRW-cutoff function~\cite{KMR}. To resolve collinear divergence
problem, we require that the modified unPDF ${\Phi}_i(x,t,\mu)$
should be satisfied exact normalization condition:
\begin{equation}
\int\limits_0^{\mu^2} dt \Phi_i(x,t,\mu^2) =
{F}_i(x,\mu^2),\label{eq:norm}
\end{equation}
or
\begin{equation}
\Phi_i(x,t,\mu^2) = \frac{d}{dt}\left[ T_i(t,\mu^2,x){F}_i(x,t)
\right],\label{eq:sudakov}
\end{equation}
where $T_i(t,\mu^2,x)$ is the Sudakov form--factor, $T_i(t =
0,\mu^2,x) = 0$ and $T_i(t = \mu^2,\mu^2,x) = 1$. The explicit form
of the Sudakov form factor in the ({\ref{eq:sudakov}) was first
obtained in~\cite{NefedovSaleev2020}:
\begin{equation}
T_i(t,\mu^2,x) = \exp\left[-\int\limits_t^{\mu^2} \frac{dt'}{t'}
\frac{\alpha_s(t')}{2\pi} \left( \tau_i(t',\mu^2) + \Delta\tau_i
(t',\mu^2,x) \right) \right],\label{eq:sud}
\end{equation}
where
\begin{eqnarray*}
\tau_i(t,\mu^2) & = & \sum\limits_j \int\limits_0^1 dz\ zP_{ji}(z)\theta
(\Delta(t,\mu^2)-z), \label{eq:tau} \\
\Delta\tau_i(t,\mu^2,x) & = & \sum\limits_j \int\limits_0^1 dz\
\theta(z - \Delta(t,\mu^2)) \left[ zP_{ji}(z) -
\frac{{F}_j\left(\frac{x}{z},t \right)}{{F}_i(x,t)} P_{ij}(z)
\theta(z - x) \right].
\end{eqnarray*}

In contrast to the KMRW model, the Sudakov form factor
(\ref{eq:sud}) depends on $x$, which is necessary to preserve the
exact normalization (\ref{eq:norm}) for any $x$ and $\mu$. The gauge
invariance of amplitudes with Reggeized partons  in the PRA
guaranteed allows you to study any processes described non-Abelian
QCD structures. PRA has been successfully used for descriptions of
angular correlations in two-jet events\cite{NSS2013}, production of
the charm \cite{Maciula:2016wci,Karpishkov:2014epa} and beauty
mesons\cite{NKS2017,Karpishkov:2016hnx}, charmonium in the NRQCD
\cite{Saleev:2012hi,He:2019qqr}.

\section{Improved Color Evaporation Model}
\label{sec:icem} The current status of the ICEM is presented in the
Ref.~\cite{ICEM_Ma_Vogt}. In the PRA, the initial partons have
transverse momenta, so the description of the spectra of single
$J/\psi$ is already possible at the leading order (LO) approximation
in the strong interaction constant in the parton subprocesses
\begin{equation}
R + R \to c + \bar{c} \label{RRcc}
\end{equation}
and
\begin{equation}
Q_q + \bar{Q}_q \to c + \bar{c},\label{QQcc}
\end{equation}
where $R$ is a Reggeized gluon, $Q_q(\bar{Q}_q)$ is a Reggeized quark
(antiquark) and $q = u, d, s$.

In the ICEM, the cross section for the production of prompt
$J/\psi$-mesons is related to the cross section for the production
of $c \bar{c}$-pairs in the single parton scattering (SPS) as
follows
\begin{eqnarray}
\sigma^{SPS}(p + p \to J/\psi + X)=\mathcal{F}^\psi \times
\int_{m_\psi}^{2m_D} \frac{d\sigma(p+p\to c+\bar
c+X)}{dM}dM,\label{eq:ICEM1}
\end{eqnarray}
where $M$ is the invariant mass of the $c \bar{c}$--pair with
4--momentum $p_{c \bar{c}}^\mu = p_c^\mu + p_{\bar{c}}^\mu$,
$m_\psi$ is the mass of the $J/\psi$--meson and $m_D$ is the mass of
the lightest $D$--meson. To take into account the kinematic effect
associated with the difference between the masses of the
intermediate state and the final charmonium, the 4--momentum of $c
\bar{c}$--pair and $J/\psi$--meson are related by $p_\psi^\mu =
(m_\psi / M) \, p_{c \bar{c}}^\mu$. The universal parameter
$\mathcal{F}^\psi$ is considered as a probability of transformation
of the $c\bar{c}$--pair with invariant mass $m_{\psi} < M < 2 m_D$
into the prompt $J/\psi$--meson.

In case of pair $J/\psi$ production via the SPS, we take into
account contributions of the following subprocesses
\begin{equation}
R+R\to c + \bar c +c + \bar c \label{RRcccc}
\end{equation}
and
\begin{equation}
Q_q+\bar Q_q \to c+ \bar c+c + \bar c.\label{QQcccc}
\end{equation}

The cross section for the production of a pair of prompt
$J/\psi$-mesons is related to the cross section for the production
of two pairs $c\bar{c}$-quarks in the following way
\begin{eqnarray}
&& \sigma^{\mathrm{SPS}}(p + p\to J/\psi + J/\psi + X) = \\
\nonumber &&=\mathcal{F}^{\psi \psi} \times \int_{m_\psi}^{2m_D}
\int_{m_\psi}^{2m_D} \frac{d\sigma(p + p \to c_1 + \bar{c}_1 + c_2 +
\bar{c}_2 + X)}{dM_1 dM_2} dM_1 dM_2,\label{eq:ICEM2}
\end{eqnarray}
where $M_{1, 2}$ are the invariant masses of $c \bar{c}$--pairs with
4--momenta $p_{c \bar{c}1}^\mu = p_{c1}^\mu + p_{\bar{c}1}^\mu$ and
$p_{c \bar{c}2}^\mu = p_{c2}^\mu + p_{\bar{c}2}^\mu$. Parameter
$\mathcal{F}^{\psi \psi}$ is the probability of transformation of two pairs
$c\bar c$ with invariant masses $m_\psi < M_{1 ,2} < 2 m_D$ into two
$J/\psi$--mesons.

In the double parton scattering (DPS) approach~\cite{DPS}, the cross
section for the production of a $J/\psi$ pair is written in terms of
the cross sections for the production of single a $J/\psi$ in two
independent subprocesses
\begin{equation}
\sigma^{\mathrm{DPS}}(p + p \to J/\psi + J/\psi + X)= \frac{
\sigma^{\mathrm{SPS}}(p + p \to J/\psi + X_1) \times
\sigma^{\mathrm{SPS}}(p + p \to J/\psi + X_2)} {2
\sigma_{\mathrm{eff}}},
\end{equation}
where the parameter $\sigma_{\mathrm{eff}}$, which controls the
contribution of the DPS mechanism is considered a free parameter.
Thus, at fitting  cross sections for pair $J/\psi$--meson
production, we assume that the parameter $\mathcal{F}^\psi$ is
fixed, and the parameters $\mathcal{F}^{\psi \psi}$ and
$\sigma_{\mathrm{eff}}$ are free parameters.

\section{Numerical methods}
\label{sec:methods}

The full gauge invariant set of Feynman diagrams of the Lipatov EFT
for the subprocess (\ref{RRcccc}) contains 72 diagrams. It is
getting too large for analytical calculation. To proceed to the next
step, we should analytically calculate squared off-shell amplitudes
and perform a numerical integration using factorization
formula~(\ref{eqI:kT_fact}) with the modified unPDFs~(\ref{uPDF}).
Nowadays, we can do it with the required numerical accuracy only for
$2\to 2$ (\ref{RRcc},\ref{QQcc}) or $2\to 3$~\cite{NKS2017}
off-shell parton subprocesses. To calculate contributions from $2
\to 4$ subprocesses with initial Reggeized partons we should apply
fully numerical methods of the calculation.

A few years ago, a new approach to obtaining gauge invariant
amplitudes with off-shell initial state partons in scattering at
high energies was proposed. The method is based on the use of spinor
amplitudes formalism and recurrence relations of the BCFW
type~\cite{hameren1,hameren2}. In Ref.~\cite{katie} was developed
the Monte Carlo (MC) parton level event generator KaTie for
processes at high energies with nonzero transverse momenta and
virtualities. This formalism~\cite{hameren1,hameren2} for numerical
amplitude generation is equivalent to amplitudes built according to
Feynman rules of the Lipatov EFT at the level of tree
diagrams~\cite{NSS2013,NKS2017,kutak}. At the stage of numerical
calculations, we use the MC event generator KaTie~\cite{katie} for
calculating the proton-proton cross sections with contributions of
all subprocesses (\ref{RRcc}), (\ref{QQcc}), (\ref{RRcccc}) and
(\ref{QQcccc}). The accuracy of numerical calculations for total
proton-proton cross sections is equal to 0.1\%.

\section{Single $J/\psi$ production}
\label{sec:results1}

We have performed the fit procedure for prompt $J/\psi$ transverse
momenta spectra in the ICEM via the PRA with  $\mathcal{F}^\psi$ as
a free parameter and obtained a rather good agreement between the
calculations and  experimental data from the energy 19.4 GeV up to
13 TeV as it was measured by different
collaborations~\cite{NA3,AFS,PHENIX,Tevatron,LHCb1a,ALICE,ATLAS,CMS,LHCb1b}.
The obtained results are collected in Table \ref{table_1} and
presented in Fig.\ref{fig_fpsi}.  Thus, as $\sqrt{s}$ decreases from
$13$ TeV to $19$ GeV, factor $F^\psi$ increases by an order of
magnitude, from about $0.02$ up to $0.2$. If we interpret the
parameter $\mathcal{F}^\psi$ as the probability of transformation of
$c\bar{c}$-pair with invariant mass from $m_\psi$ to $2m_D$ into
$J/\psi$-meson, its growth with decreasing energy can be explained
by an increase of the hadronization time. Energy dependence of the
$\mathcal{F}^\psi$ well described by a formula
\begin{equation}
\mathcal{F}^\psi(\sqrt{s}) = 0.012 + 0.952
(\sqrt{s})^{-0.525}.\label{eq:fpsi}
\end{equation}
The calculated transverse momentum spectra and the experimental data
are presented in Figs. \ref{fig_LHCball})-(\ref{fig_NA3}. Grey boxes
around the central lines in the Figures indicate upper and lower
limits of the cross-section obtained due to variation of the hard
scale $\mu$ by the factors $\xi=2$ or $\xi=1/2$ around the central
value of the hard scale $(\mu=\sqrt{m_\psi^2+p_T^2})$ and the
$c$-quark mass from $1.2$ to $1.4$ GeV.

In contrast to the predictions obtained in the NRQCD approach, when
gluon-gluon fusion in the $J/\psi$ hadroproduction is the dominant
mechanism,  the ICEM predicts a sufficiently large contribution from
the process of quark-antiquark annihilation especially at low
energy, see Fig. \ref{fig_ratio}. Thus, at the energy of future
proton-proton collider NICA, $\sqrt{s}\simeq 30$ GeV, the
quark-antiquark contribution may be about 30 \% of the total cross
section of prompt $J/\psi$ production.
\begin{figure}[h]
\begin{center}
    \includegraphics[width=0.8\textwidth]{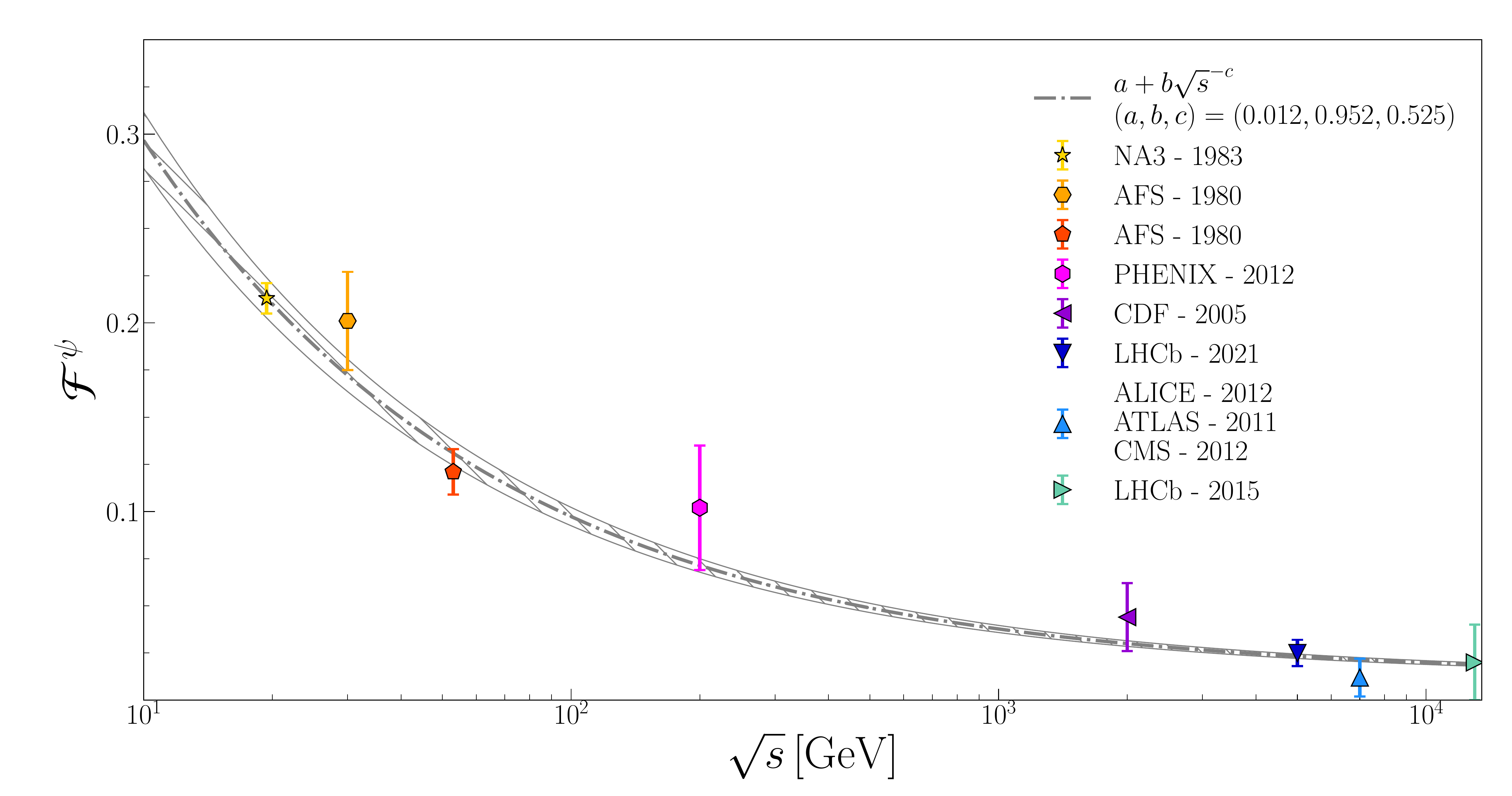}
\vspace{-3mm} \caption{ The hadronization parameter
$\mathcal{F}^{\psi}$ as a function of proton collision energy
$\sqrt{s}$. The corridor between the upper and lower lines
demonstrates the uncertainty from the hard scale variation by the
factor $\xi = 2$  and the $c$-quark mass from $1.2$ to $1.4$ GeV.
\label{fig_fpsi}}
\end{center}
\end{figure}

\small
\begin{table}[p]
\caption{The parameter ${\mathcal F}^\psi$ at the kinematical
conditions of the different experiments}

\begin{center}
\begin{tabular}{l c c c r}
\hline \hline
Collaboration & Energy & Rapidity & Transverse momentum & $\mathcal{F}^{\psi}$ \\
\hline

NA3: & $\sqrt{s} = 19.4$ GeV & $y > 0$ & $p_T \in [0, 5]$ GeV &
$0.213^{+0.008}_{-0.008}$ \\
\hline

AFS: & $\sqrt{s} = 30$ GeV & $|y| < 0.5$ & $p_T \in [0, 5]$ GeV &
$0.201^{+0.026}_{-0.026}$ \\
\hline
& $\sqrt{s} = 53$ GeV & $|y| < 0.5$ & $p_T \in [0, 7]$ GeV & $0.121^{+0.012}_{-0.012}$ \\
\hline

PHENIX: & $\sqrt{s} = 200$ GeV & $|y| < 0.35$ & $p_T \in [0, 9]$ GeV
&
$0.102^{+0.033}_{-0.033}$ \\
\hline

CDF: & $\sqrt{s} = 1.96$ TeV & $|y| < 0.6$ & $p_T \in [0, 20]$ GeV &
$0.044^{+0.018}_{-0.018}$ \\
\hline

LHCb: & $\sqrt{s} = 5$ TeV & $2.0 < y < 2.5$ & $p_T \in [0, 14]$ GeV
&
$0.025^{+0.007}_{-0.007}$ \\
\hline

ALICE: & $\sqrt{s} = 7$ TeV & $|y| < 0.9$ & $p_T \in [0, 7]$ GeV &
$0.037^{+0.007}_{-0.007}$ \\
\hline

ATLAS: & $\sqrt{s} = 7$ TeV & $|y| < 0.75$ & $p_T \in [7, 70]$ GeV &
$0.013^{+0.002}_{-0.001}$ \\
\hline & & $0.75 < |y| < 1.50$ & $p_T \in [5, 70]$ GeV &
$0.007^{+0.001}_{-0.001}$ \\
\hline & & $1.5 < |y| < 2.0$ & $p_T \in [1, 30]$ GeV &
$0.011^{+0.001}_{-0.001}$ \\
\hline & & $2.0 < |y| < 2.4$ & $p_T \in [5, 30]$ GeV &
$0.009^{+0.001}_{-0.001}$ \\
\hline

CMS:& $\sqrt{s} = 7$ TeV & $|y| < 0.9$ & $p_T \in [8, 70]$ GeV &
$0.005^{+0.001}_{-0.001}$ \\
\hline & & $0.9 < |y| < 1.2$ & $p_T \in [8, 45]$ GeV &
$0.007^{+0.001}_{-0.002}$ \\
\hline & & $1.2 < |y| < 1.6$ & $p_T \in [6.5, 45]$ GeV &
$0.007^{+0.002}_{-0.002}$ \\
\hline & & $1.6 < |y| < 2.1$ & $p_T \in [6.5, 30]$ GeV &
$0.009^{+0.002}_{-0.002}$ \\
\hline & & $2.1 < |y| < 2.4$ & $p_T \in [5.5, 30]$ GeV &
$0.009^{+0.002}_{-0.002}$ \\
\hline

LHCb: & $\sqrt{s} = 13$ TeV & $2.0 < y < 2.5$ & $p_T \in [0, 14]$
GeV &
$0.021^{+0.004}_{-0.004}$ \\
\hline & & $2.5 < y < 3.0$ & $p_T \in [0, 14]$ GeV &
$0.022^{+0.004}_{-0.004}$ \\
\hline & & $3.0 < y < 3.5$ & $p_T \in [0, 14]$ GeV &
$0.021^{+0.004}_{-0.004}$ \\
\hline & & $3.5 < y < 4.0$ & $p_T \in [0, 14]$ GeV &
$0.018^{+0.005}_{-0.005}$ \\
\hline \hline \label{table_1}
\end{tabular}

\end{center}

\end{table}

\normalsize

\begin{figure}[h]
\begin{center}
\vspace{0cm}
  \includegraphics[width=0.8\textwidth,angle=0]{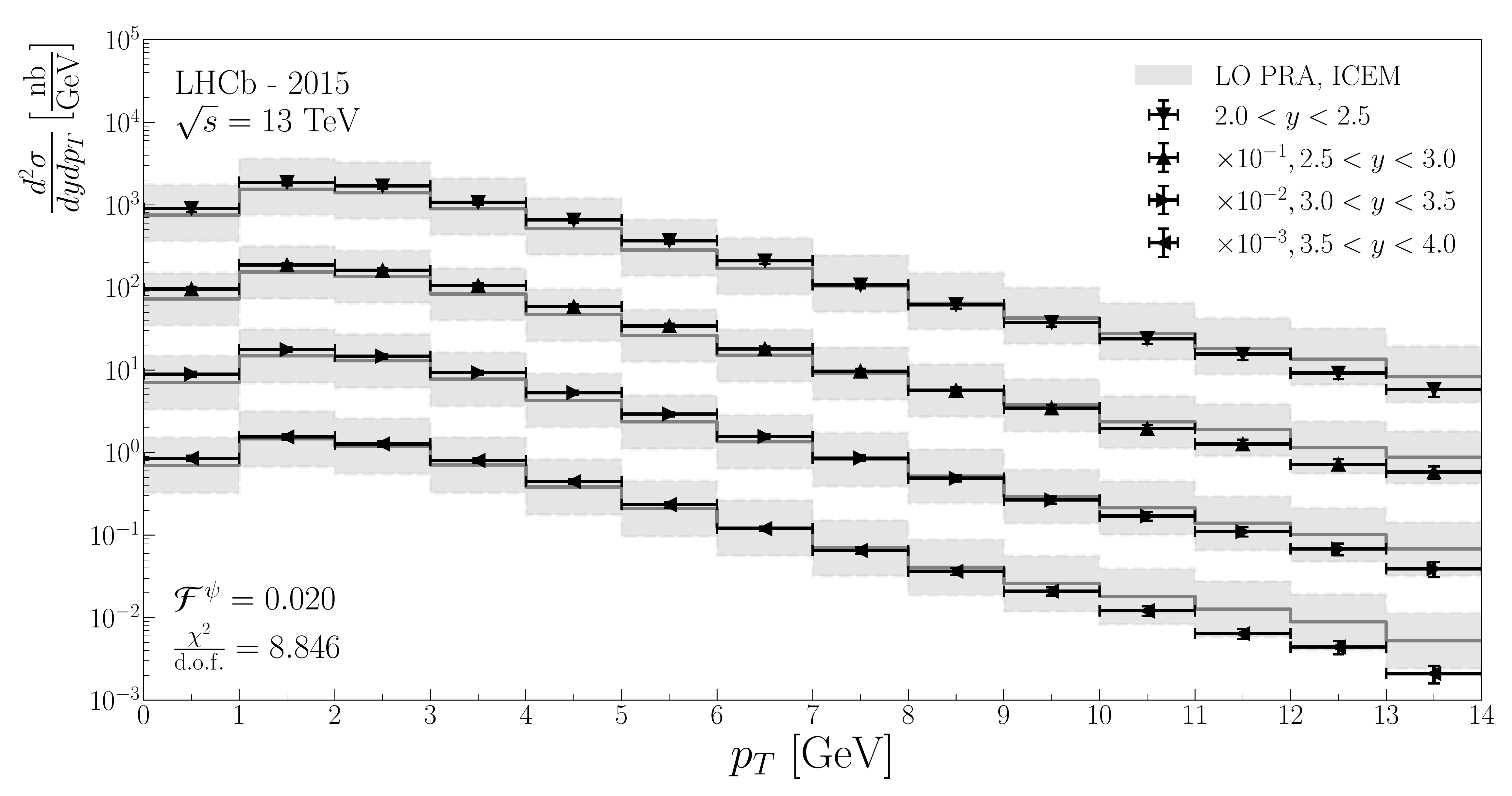}
    \end{center}
\caption{ The transverse momentum spectra of prompt $J/\psi$ at the
different ranges of rapidities. The data are from LHCb collaboration
at the $\sqrt{s}=13$ TeV \cite{LHCb1b}. \label{fig_LHCball}}
\end{figure}

\begin{figure}[h]
\begin{center}
\vspace{0cm}
  \includegraphics[width=0.8\textwidth,angle=0]{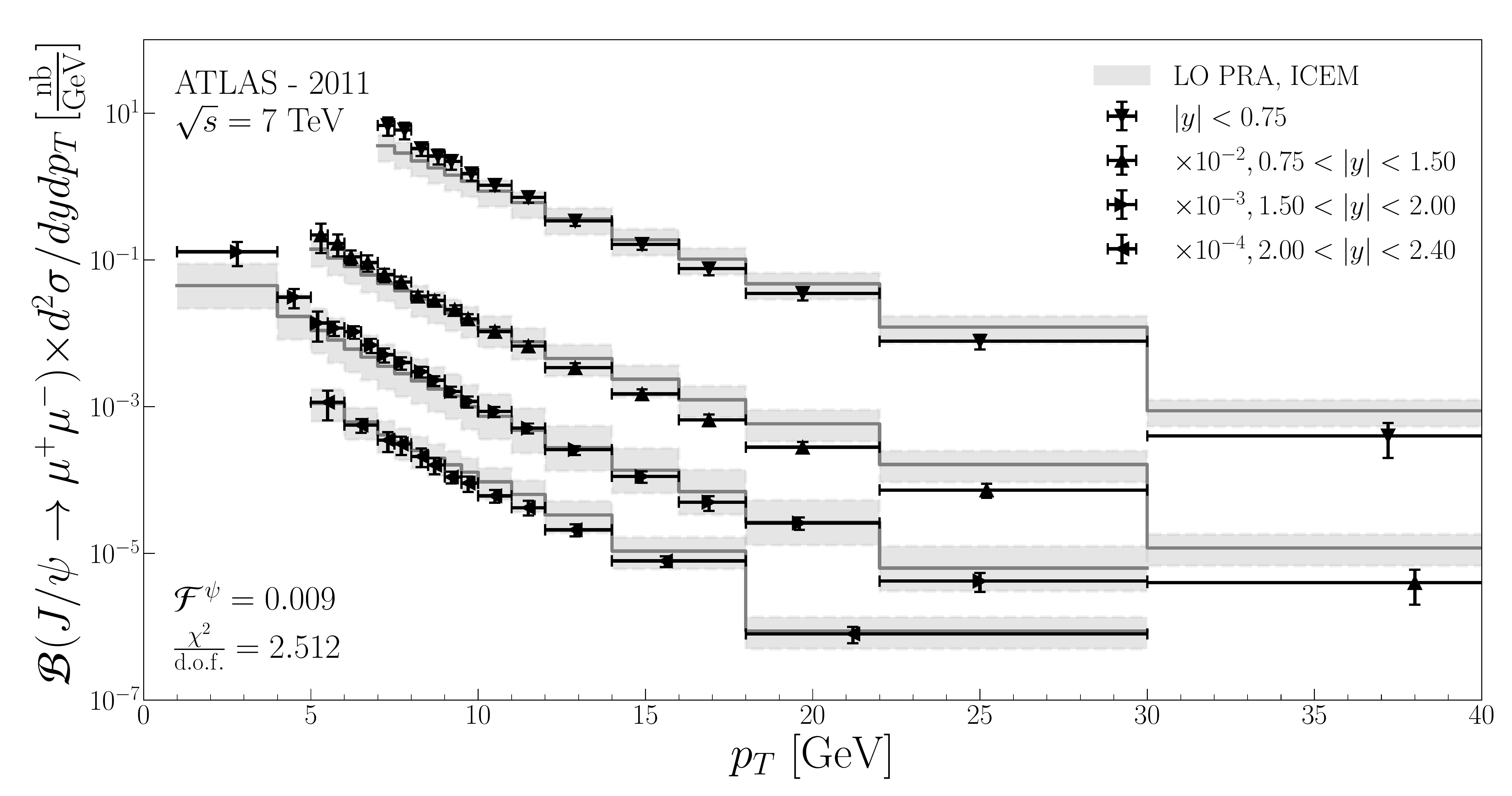}
    \end{center}
\caption{ The transverse momentum spectra of prompt $J/\psi$ at the
different ranges of rapidities. The data are from ATLAS
collaboration at the $\sqrt{s}=8$ TeV \cite{ATLAS}.
\label{fig_ATLASall}}
\end{figure}

\begin{figure}[h]
\begin{center}
\vspace{0cm}
  \includegraphics[width=0.8\textwidth,angle=0]{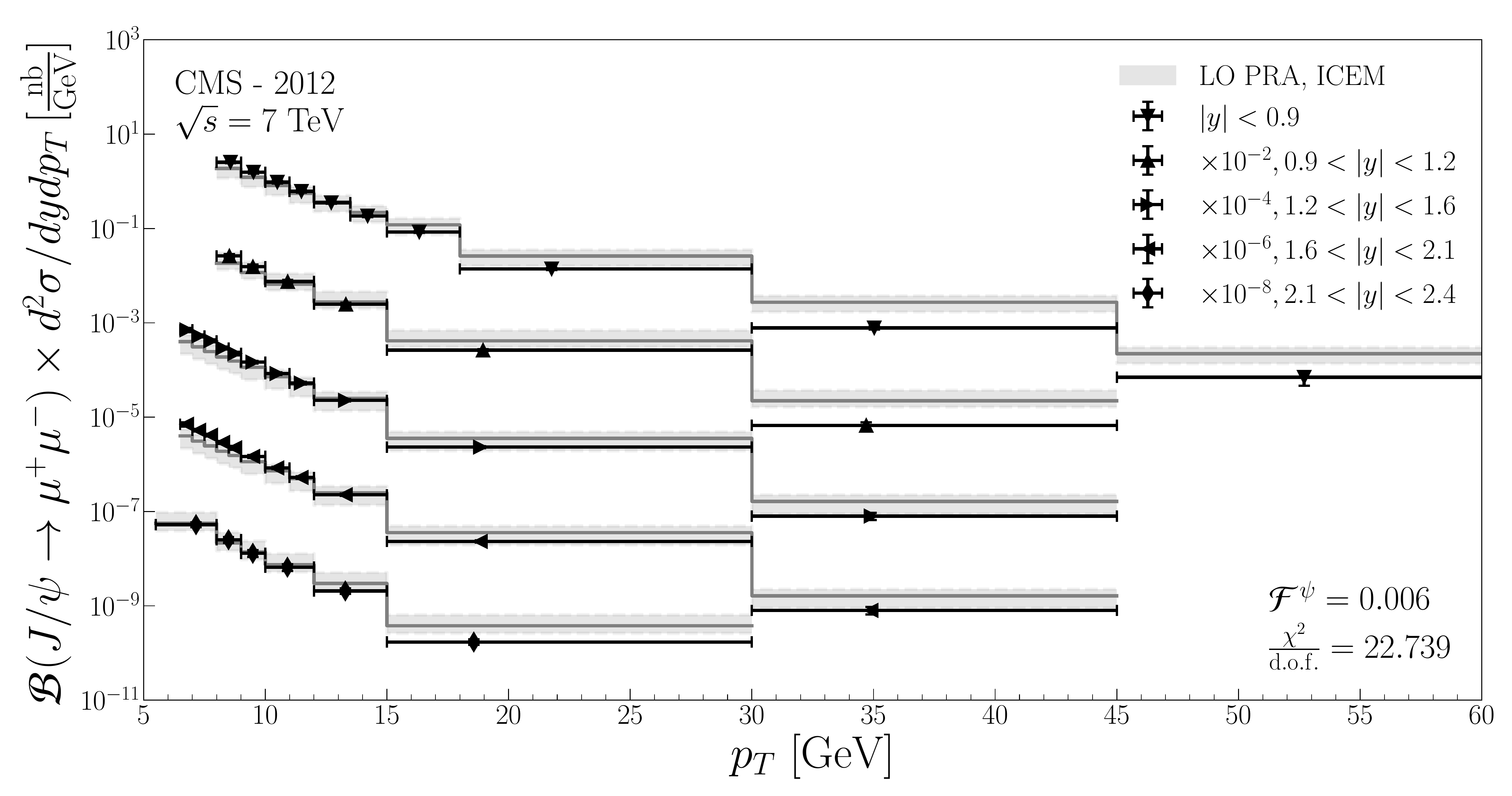}
    \end{center}
\caption{ The transverse momentum spectra of prompt $J/\psi$ at the
different ranges of rapidities. The data are from CMS collaboration
at the $\sqrt{s}=8$ TeV \cite{CMS}. \label{fig_CMSall}}
\end{figure}

\begin{figure}[h]
\begin{center}
\vspace{0cm}
  \includegraphics[width=0.45\textwidth,angle=0]{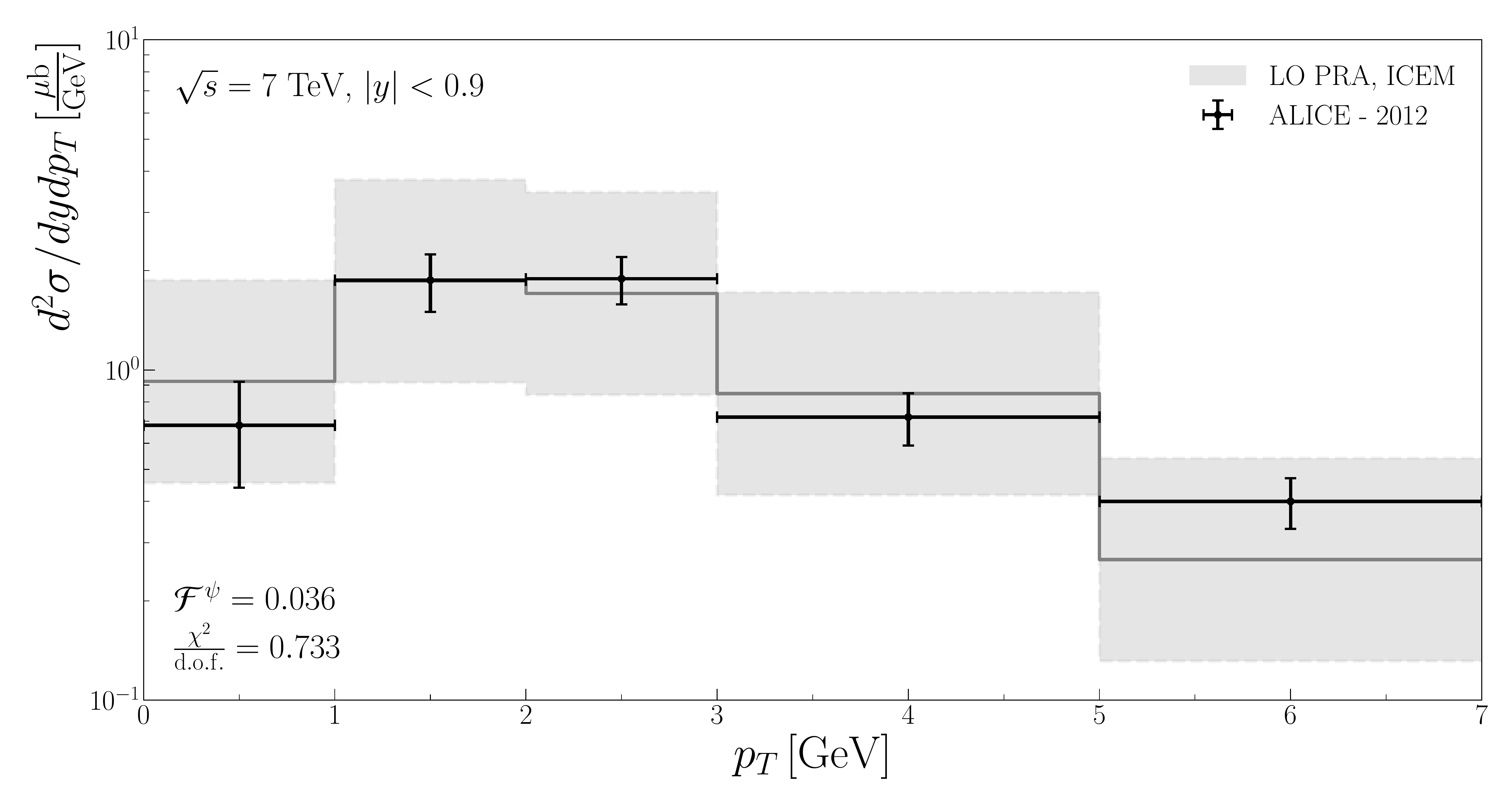}  \includegraphics[width=0.45\textwidth,angle=0]{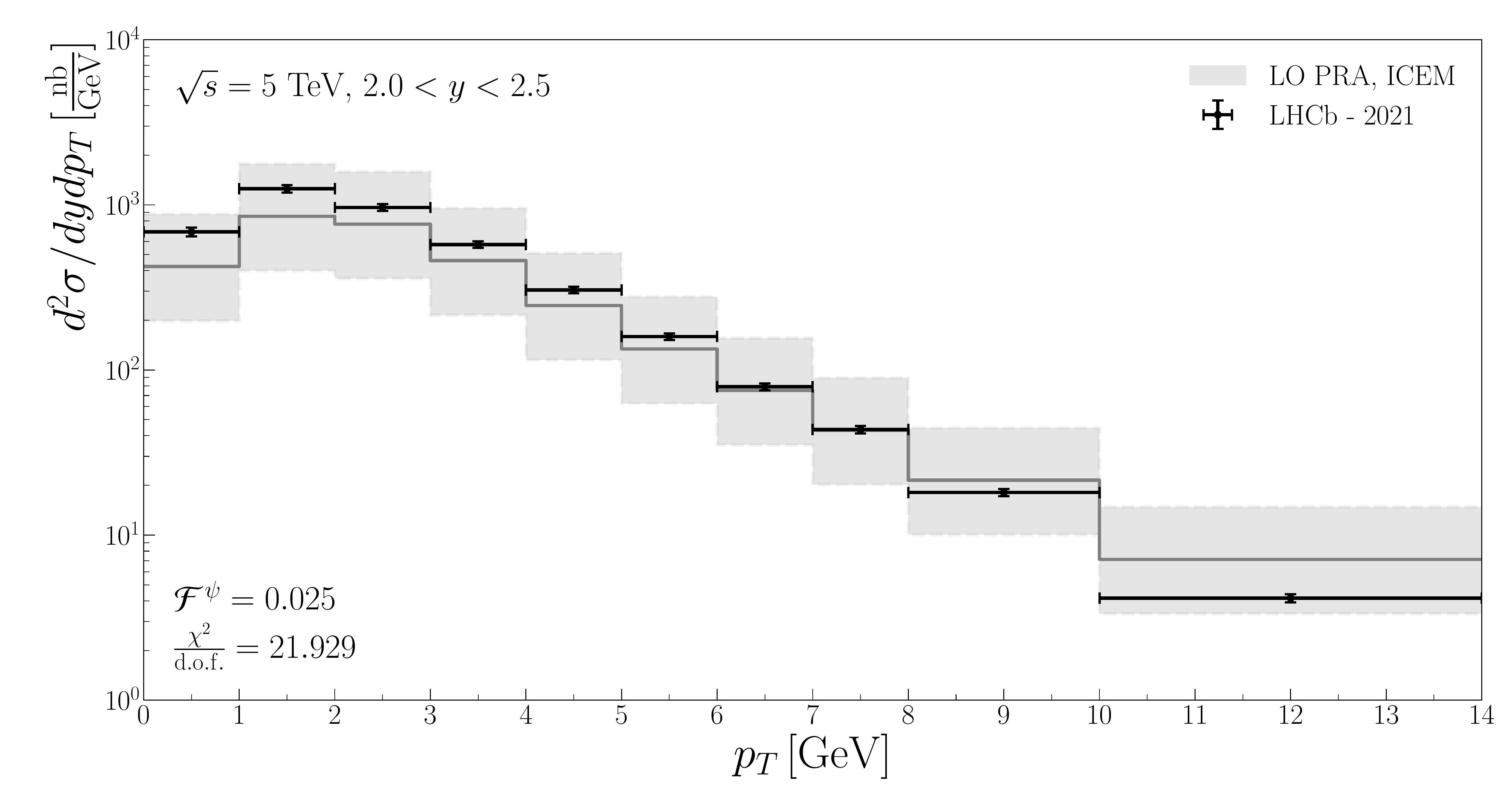}
    \end{center}
\caption{ The transverse momentum spectra of prompt $J/\psi$. In the
left panel, the data are from ALICE collaboration at the
$\sqrt{s}=7$ TeV \cite{ALICE}. In the right panel, the data are from
LHCb collaboration at the $\sqrt{s}=5$ TeV \cite{LHCb1a}.
\label{fig_ALICE}}
\end{figure}

\begin{figure}[h]
\begin{center}
\vspace{0cm}
  \includegraphics[width=0.45\textwidth,angle=0]{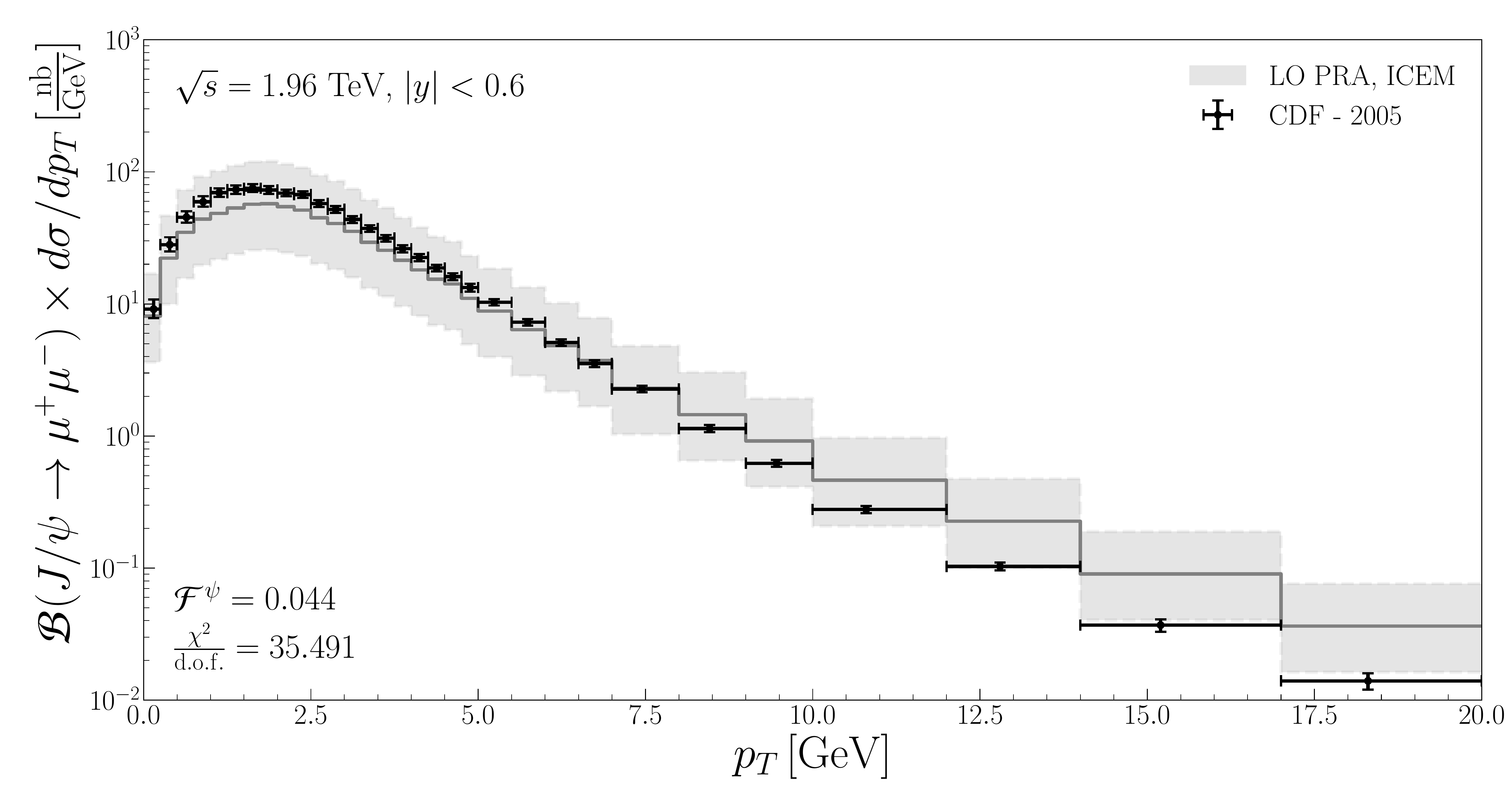}\includegraphics[width=0.45\textwidth,angle=0]{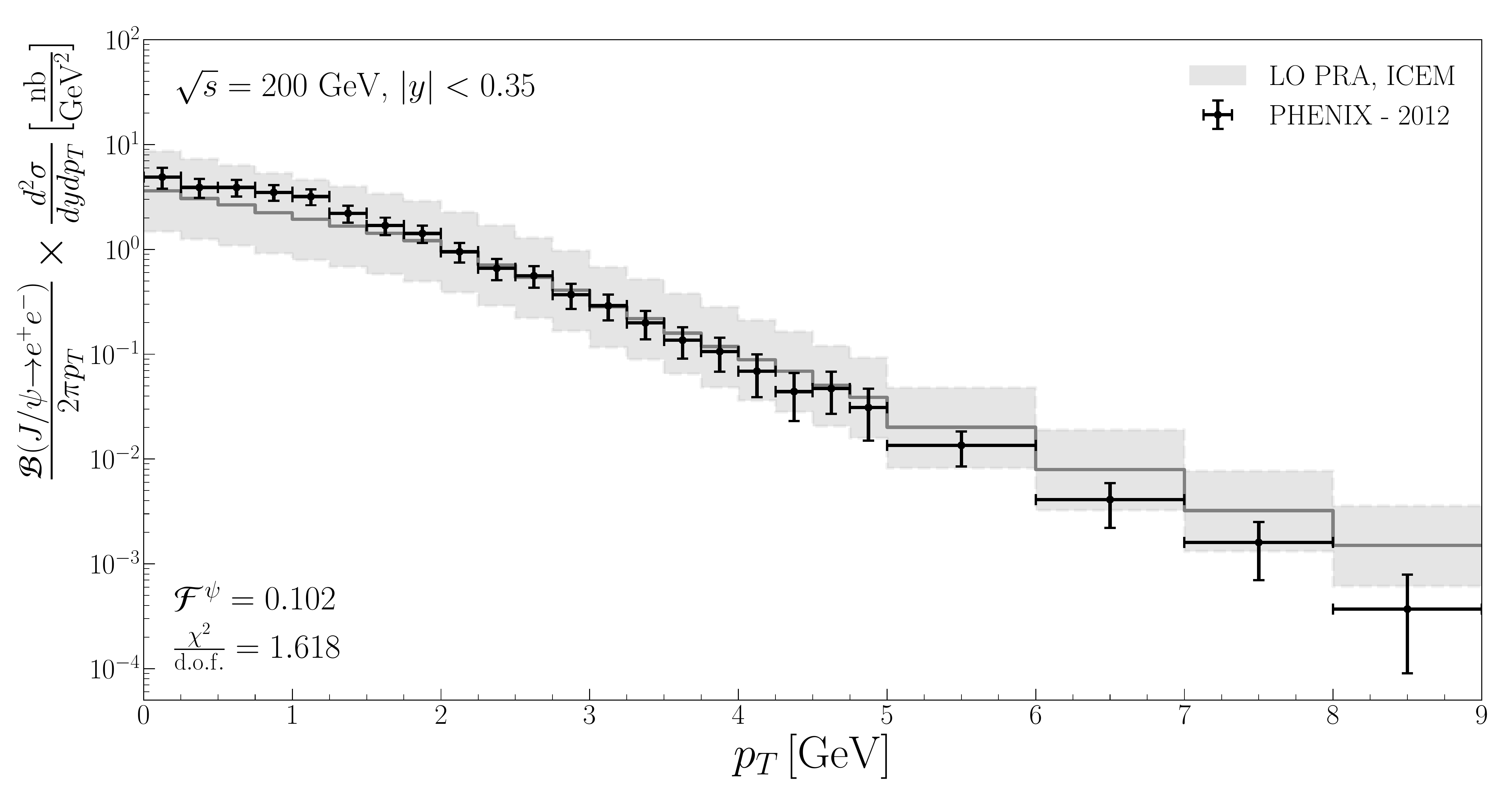}
    \end{center}
\caption{ The transverse momentum spectra of prompt $J/\psi$. In the
left panel, the data are from CDF collaboration at the
$\sqrt{s}=1.8$ TeV \cite{Tevatron}. In the right panel, the data are
from PHENIX collaboration at the $\sqrt{s}=0.2$ TeV \cite{PHENIX}.
\label{fig_PHENIX}}
\end{figure}

\begin{figure}[h]
\begin{center}
\vspace{0cm}
  \includegraphics[width=0.45\textwidth,angle=0]{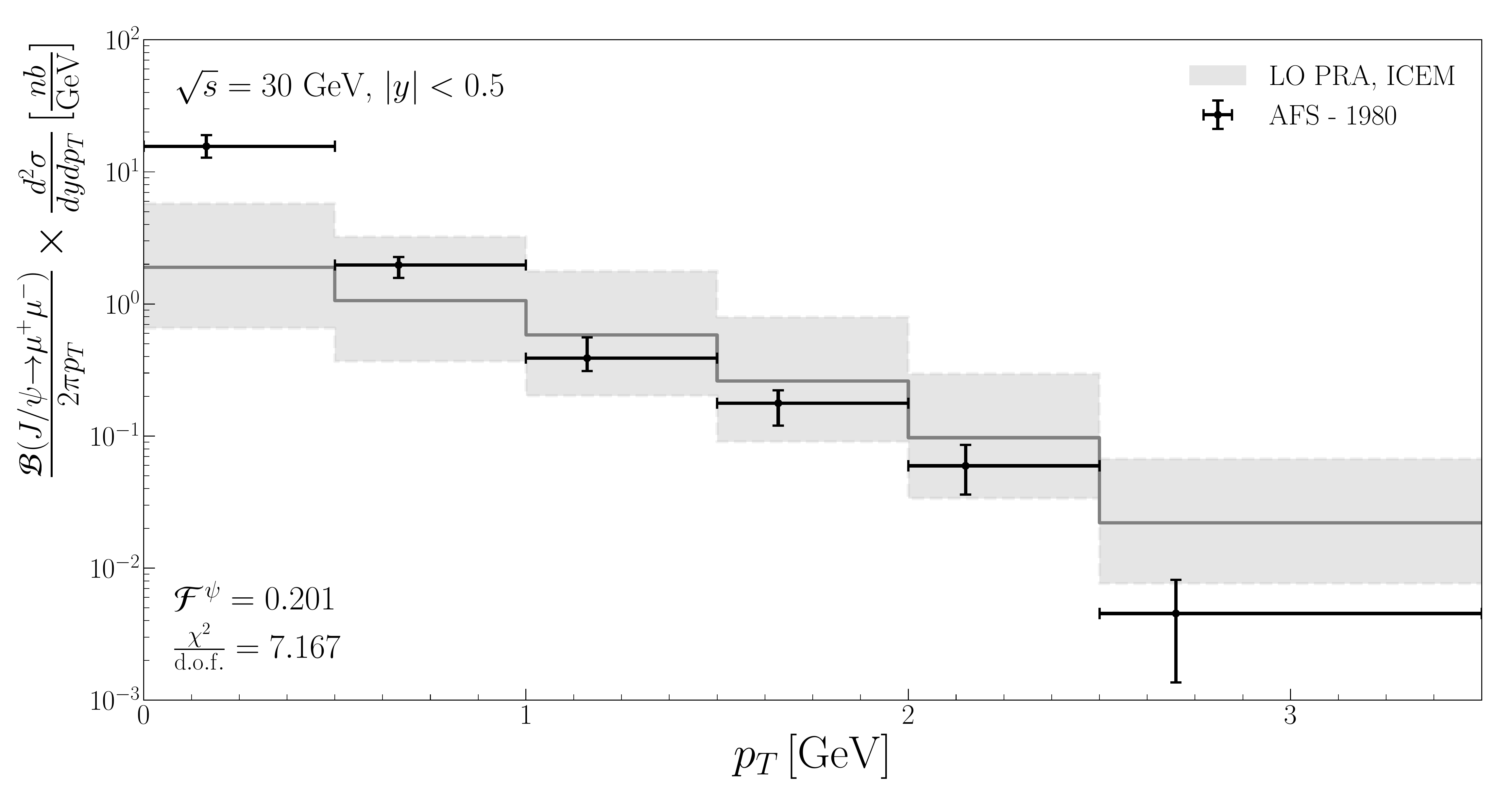}\includegraphics[width=0.45\textwidth,angle=0]{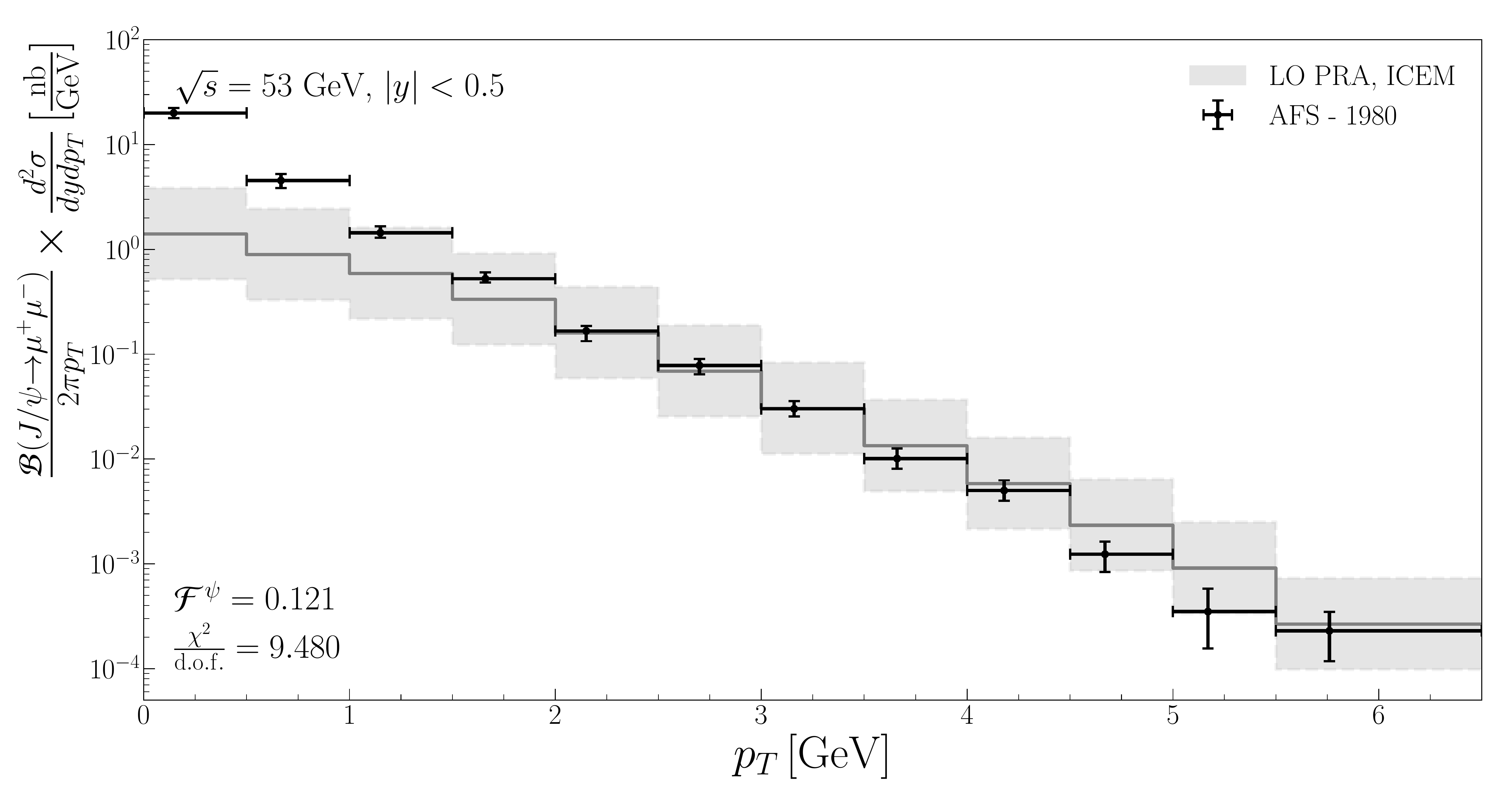}
    \end{center}
\caption{ The transverse momentum spectra of prompt $J/\psi$. In the
left panel, the data are from AFS collaboration, at the
$\sqrt{s}=30$ GeV \cite{AFS}. In the right panel, the data are from
AFS collaboration at the $\sqrt{s}=53$ GeV \cite{AFS}.
\label{fig_ASF}}
\end{figure}

\begin{figure}[h]
\begin{center}
\vspace{0cm}
  \includegraphics[width=0.8\textwidth,angle=0]{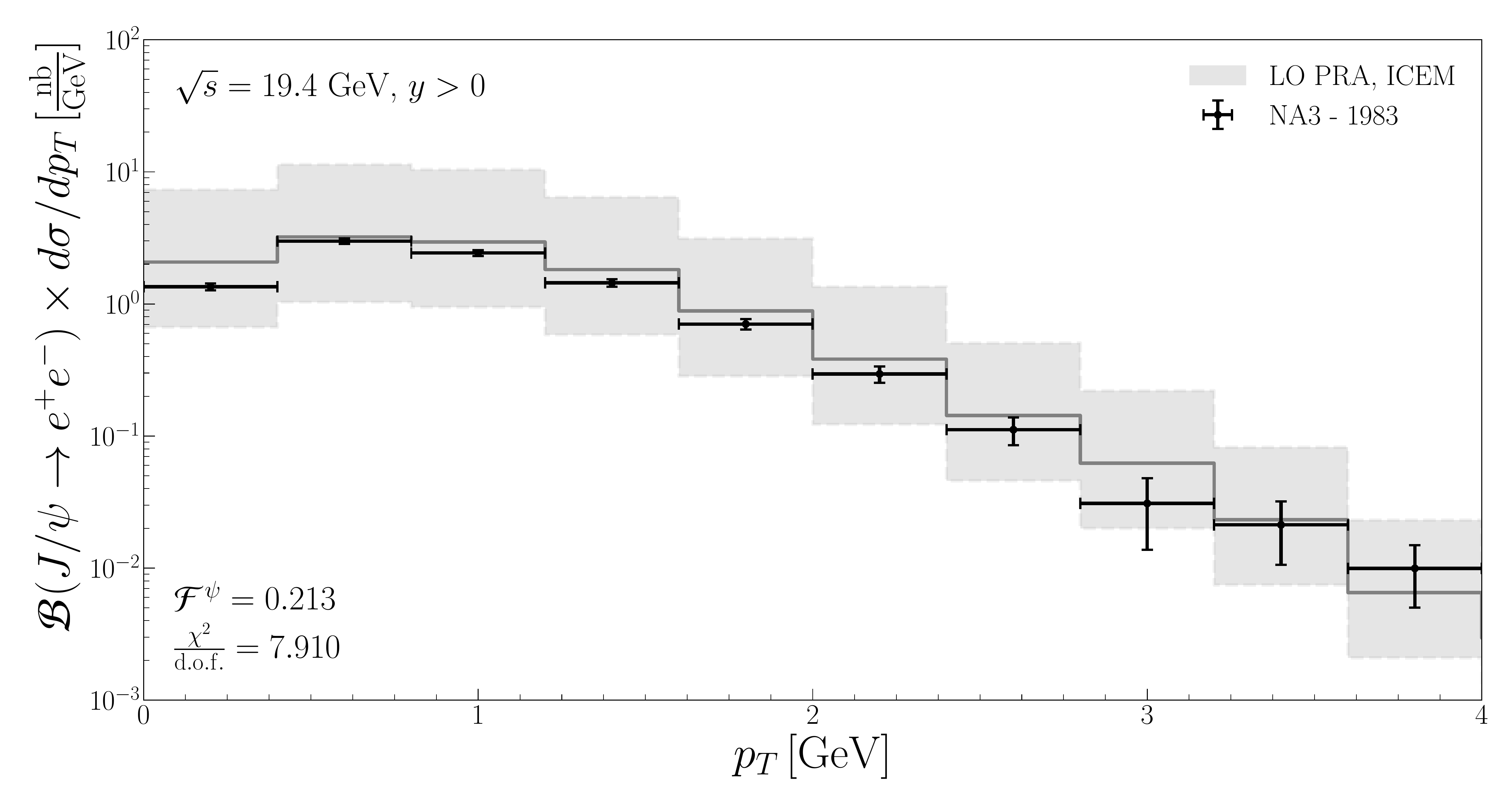}
    \end{center}
\caption{ The transverse momentum spectrum of prompt $J/\psi$. The
data are from NA3 collaboration at the $\sqrt{s}=19.4$ GeV
\cite{NA3}. \label{fig_NA3}}
\end{figure}

\begin{figure}[h]
\begin{center}
    \includegraphics[width=0.8\textwidth]{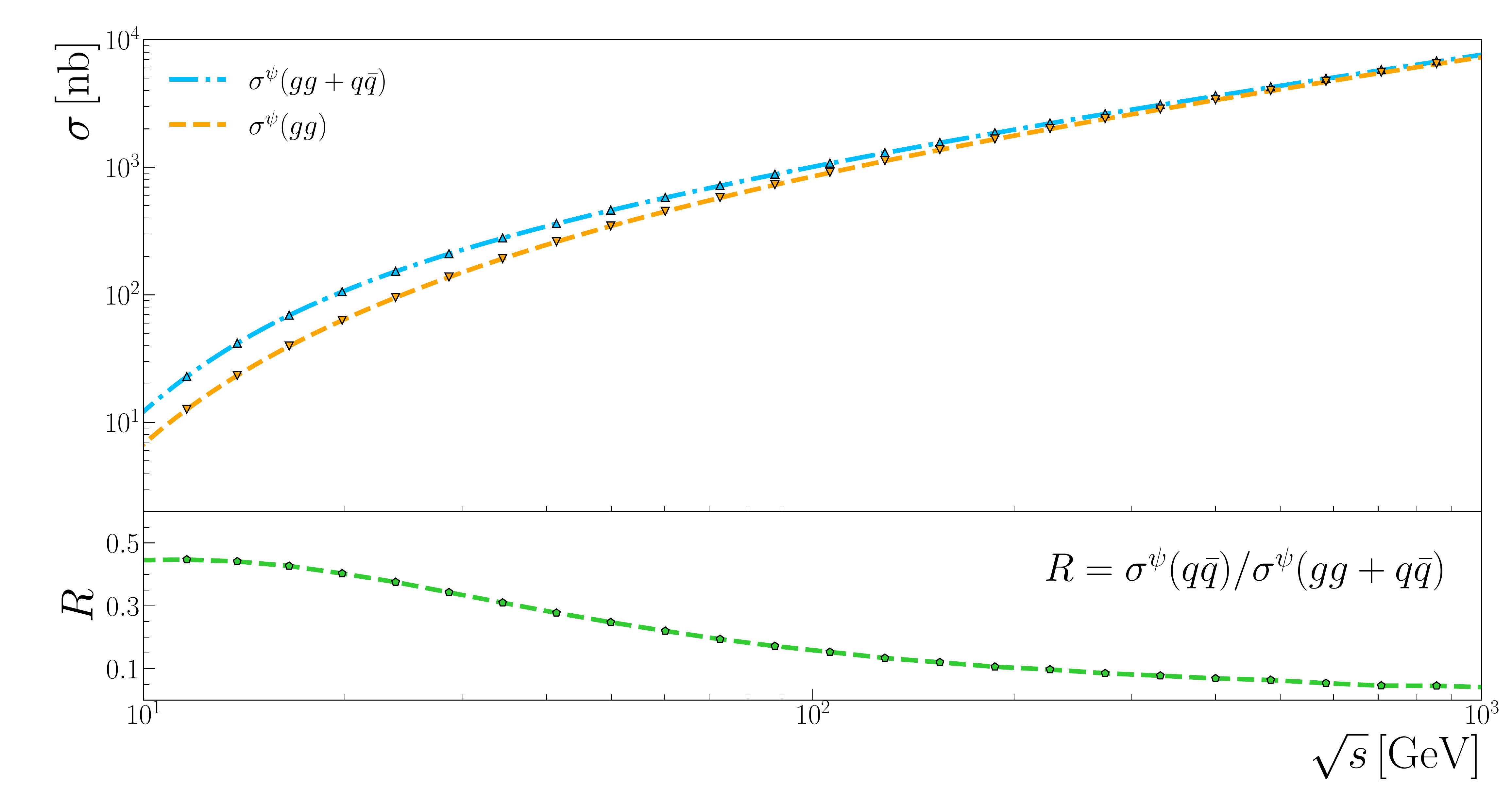}
\vspace{-3mm} \caption{The relative contributions of the parton
subprocess (\ref{RRcc}) and (\ref{QQcc}) in the prompt $J/\psi$
production as a function of energy obtained in the ICEM via the PRA.
\label{fig_ratio}}
\end{center}
\end{figure}

\clearpage
\section{Pair $J/\psi$ production}
\label{sec:results2}

In Ref.~\cite{lansberg}, the pair $J/\psi$ production was studied in
the next to leading order approximation of the collinear parton
model. The authors assumed
\begin{equation}
{\mathcal{F}}^{\psi\psi} = \left({\mathcal{F}}^{\psi}\right)^2 \label{eq:f2psi}
\end{equation}
and found that the contribution of the SPS production mechanism is
negligible and the experimental data can only be described by the
DPS mechanism. In our opinion, the relation (\ref{eq:f2psi}) is
valid only in the case of the dominant role of the fragmentation
approximation for the production of the $J/\psi$ pair. However, the
fragmentation mechanism of $J/\psi$ production becomes dominant for
$p^{\psi}_{T} \geq 15$ GeV, i.e. at much larger transverse momentum
of the $J/\psi$ than at which the measurements
\cite{CMS_2Psi,ATLAS_2Psi,LHCb_2Psi} were made.

First of all, we review the setup of the pair $J/\psi$ measurements:
\begin{itemize}
\item LHCb, $\sqrt{s} = 13$ TeV , $2.0 < y^{\psi} < 4.5$ , $p^{\psi}_T <
10.0$ GeV.
\item ATLAS (REG-I), $\sqrt{s} = 8$ TeV , $|y^{\psi_1}| < 2.10$ ,
 $p_T^{\psi_1} >
8.5$ GeV , $|y^{\psi_2}| < 1.05$ , $p_T^{\psi_2} > 8.5$ GeV, where
 $p_T^{\psi_2} < p_T^{\psi_1}$.
\item ATLAS (REG-II),  $\sqrt{s} = 8$ TeV , $|y^{\psi_1}| < 2.10$ ,
$p_T^{\psi_1} > 8.5$ GeV , $1.05 < |y^{\psi_2}| < 2.10$ ,
$p_T^{\psi_2} > 8.5$ GeV, where $p_T^{\psi_2} < p_T^{\psi_1}$.
\item CMS (REG-I), $\sqrt{s} = 7$ TeV , $|y^{\psi}| < 1.20$ , $p_T^{\psi} > 6.5$
GeV.
\item CMS (REG-II), $\sqrt{s} = 7$ TeV , $1.20 < |y^{\psi}| < 1.43$ , $p_T^{\psi} \in (6.5 \to 4.5)$
GeV.
\item CMS (REG-III), $\sqrt{s} = 7$ TeV , $1.43 < |y^{\psi}| < 2.20$ , $p_T^{\psi} > 4.5$
GeV.
\end{itemize}





In case of pair $J/\psi$ production, we took into account the
contributions of the SPS and the DPS production mechanisms.
Parameters $\mathcal{F}^{\psi \psi}$ and $\sigma_{\mathrm{eff}}$
obtained by the separate fits of production cross sections for
different experiments are shown as a contour plot in the
Fig.\ref{fig_fpsi2a}. Two curves for each experiment
($k=LHCb,ATLAS,CMS$) correspond $x_k=\pm 1$, where
$$
x_k = \frac{\sigma^{\mathrm{exp}}_k - \sigma^{\mathrm{theor}}_k}
{\Delta\sigma^{\mathrm{exp}}_k}.
$$
We find that there is a common region of parameters
$\mathcal{F}^{\psi \psi}$ and $\sigma_{\mathrm{eff}}$ for all
experiments. If we collect all experimental data into one set for
fit, we find more strong conditions in a plane of these two
parameters, which are shown as a contour plot in
Fig.\ref{fig_fpsi2b}. The isolines correspond to the numerical
values of the parameter $x = 1.0, 1.5$ and $2.0$, where
$$
x = \sum_{k = 1}^n \frac{|\sigma^{\mathrm{exp}}_k -
\sigma^{\mathrm{theor}}_k|} {\Delta\sigma^{\mathrm{exp}}_k}
$$
and the sum is taken over all cross-sections of three experiments:
CMS~\cite{CMS_2Psi}, ATLAS~\cite{ATLAS_2Psi} and
LHCb~\cite{LHCb_2Psi}. The best description of the data, when $x <
1.0$, is reached in the parameter domain $0.021 < \mathcal{F}^{\psi
\psi} < 0.023$ and $10.75 < \sigma_{\mathrm{eff}} < 11.2$ mb. In
fact, at the LHC energies one has $\mathcal{F}^{\psi\psi} \simeq
\mathcal{F}^{\psi}$. The optimal obtained value for
$\sigma_{\mathrm{eff}}$ is in a good agreement with the estimates
obtained early in other studies~\cite{CMS_2Psi,LHCb_2Psi}.

\begin{figure}[t]
\begin{center}
    \includegraphics[width=0.8\textwidth]{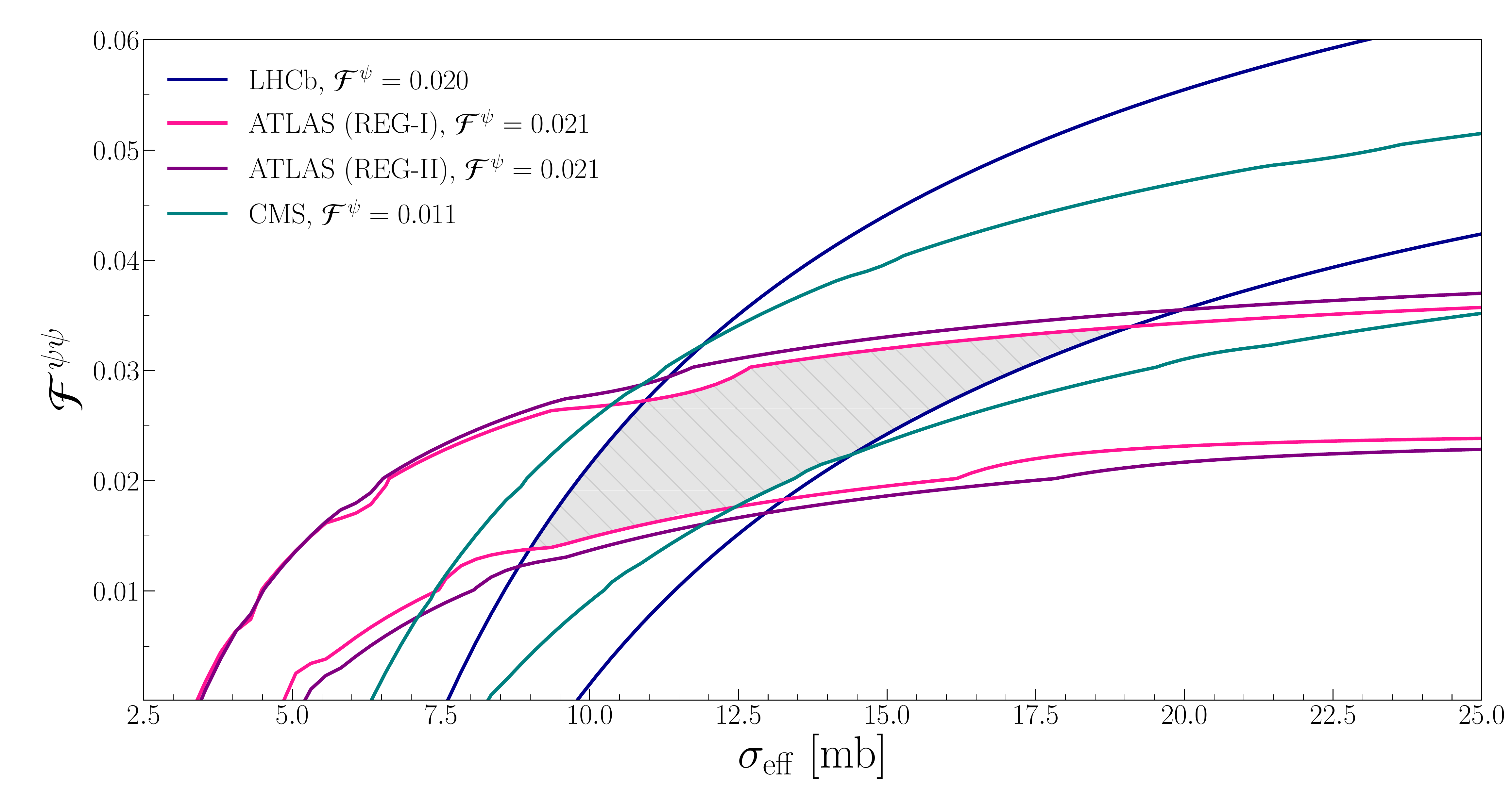}
\vspace{-3mm} \caption{Regions of the parameters $\mathcal{F}^{\psi
\psi}$ and $\sigma_{\mathrm{eff}}$ in the ICEM for pair $J/\psi$
production, obtained as a result of data fitting. The relevant pairs
of isolines correspond to $x_k = \pm 1.0$ for different experiments.
\label{fig_fpsi2a}}
\end{center}
\vspace{-5mm}
\end{figure}

\begin{figure}[t]
\begin{center}
    \includegraphics[width=0.8\textwidth]{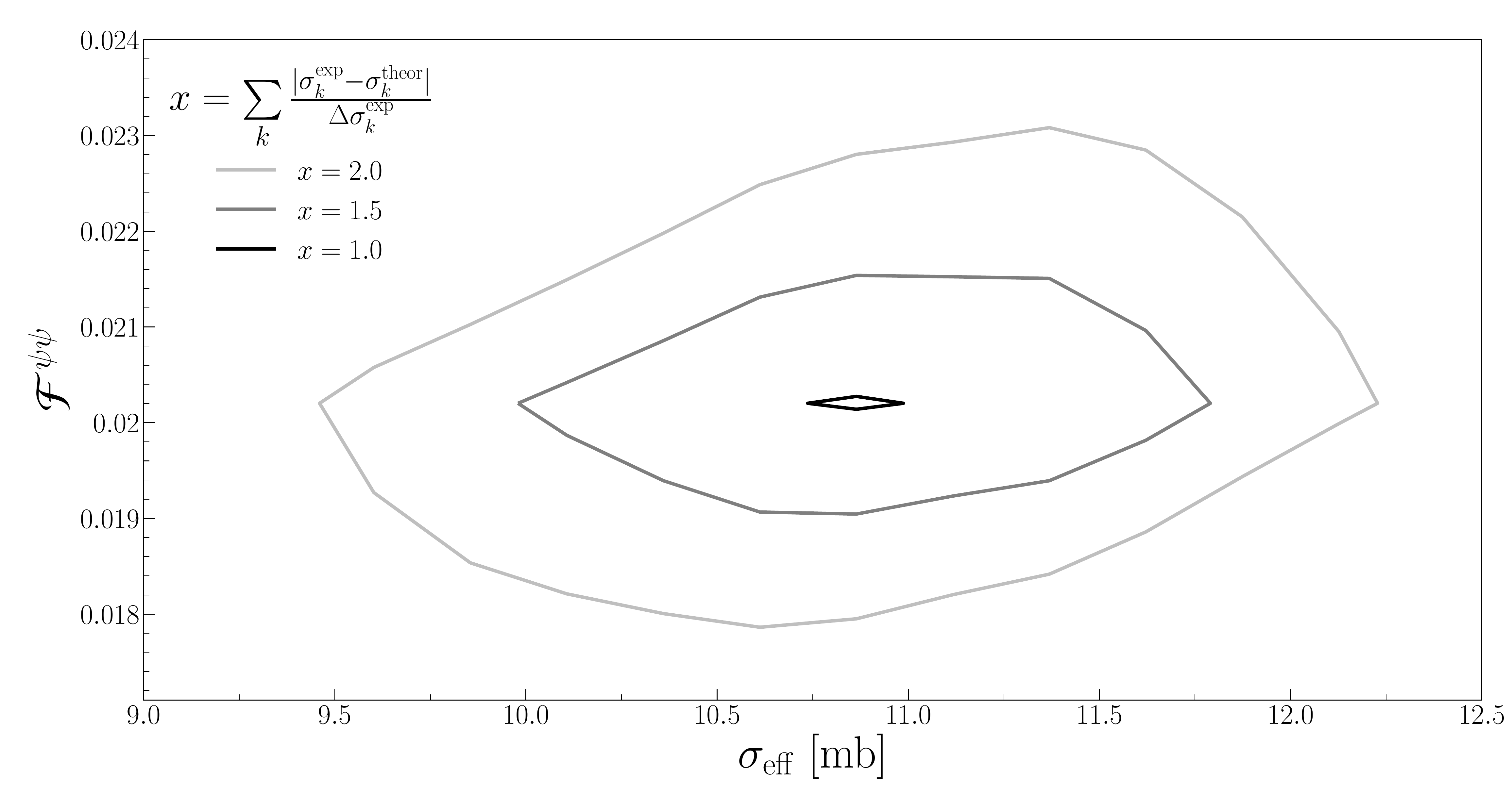}
\vspace{-3mm} \caption{Regions of the parameters $\mathcal{F}^{\psi
\psi}$ and $\sigma_{\mathrm{eff}}$ in the ICEM for pair $J/\psi$
production, obtained as a result of data fitting. Isolines
correspond to $x = 1.0, 1.5$ and $2.0$. \label{fig_fpsi2b}}
\end{center}
\vspace{-5mm}
\end{figure}

To demonstrate agreement between our calculations in the ICEM via
the PRA and experimental data for pair $J/\psi$ production, we plot
in the Figs. \ref{fig_LHCb2})-(\ref{fig_CMS2} different spectra,
which have been obtained with $\mathcal{F}^{\psi} =0.02$,
$\mathcal{F}^{\psi \psi} = 0.02$ and $ \sigma_{\mathrm{eff}} = 11$
mb. It is interesting compare the ratio of the SPS and DPS
contributions, $R=\sigma^{SPS}_{\psi\psi}/\sigma^{DPS}_{\psi\psi}$,
to the pair $J/\psi$ production cross sections with the above
mentioned values of $\mathcal{F}^{\psi}$, $\mathcal{F}^{\psi \psi}$
and $ \sigma_{\mathrm{eff}}$: for the LHCb data ($\sqrt{s}=13$ TeV)
-- $R\simeq 0.2$, for the  CMS data ($\sqrt{s}=7$ TeV) -- $R\simeq
0.5$, but for the ATLAS data ($\sqrt{s}=8$ TeV) -- $R\simeq 1.5$ .
In such a way, the DPS production mechanism is a dominant source of
$J/\psi$ pairs only when the both $J/\psi$ are produced in the
forward region of rapidity, as it is measured by LHCb Collaboration.

\begin{figure}[h]
\begin{center}
\vspace{0cm}
\includegraphics[width=0.45\textwidth,angle=0]{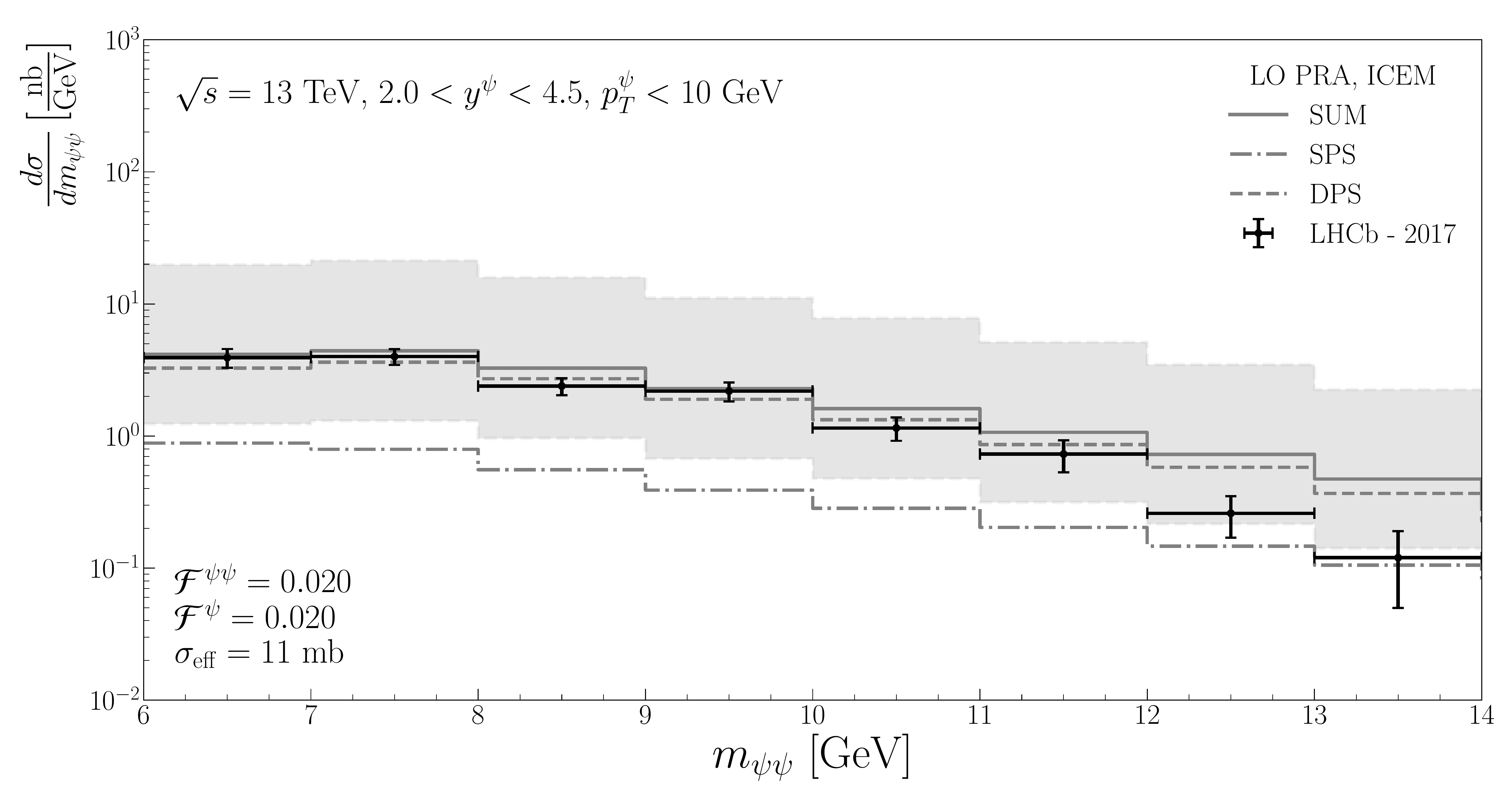}\includegraphics[width=0.45\textwidth,angle=0]{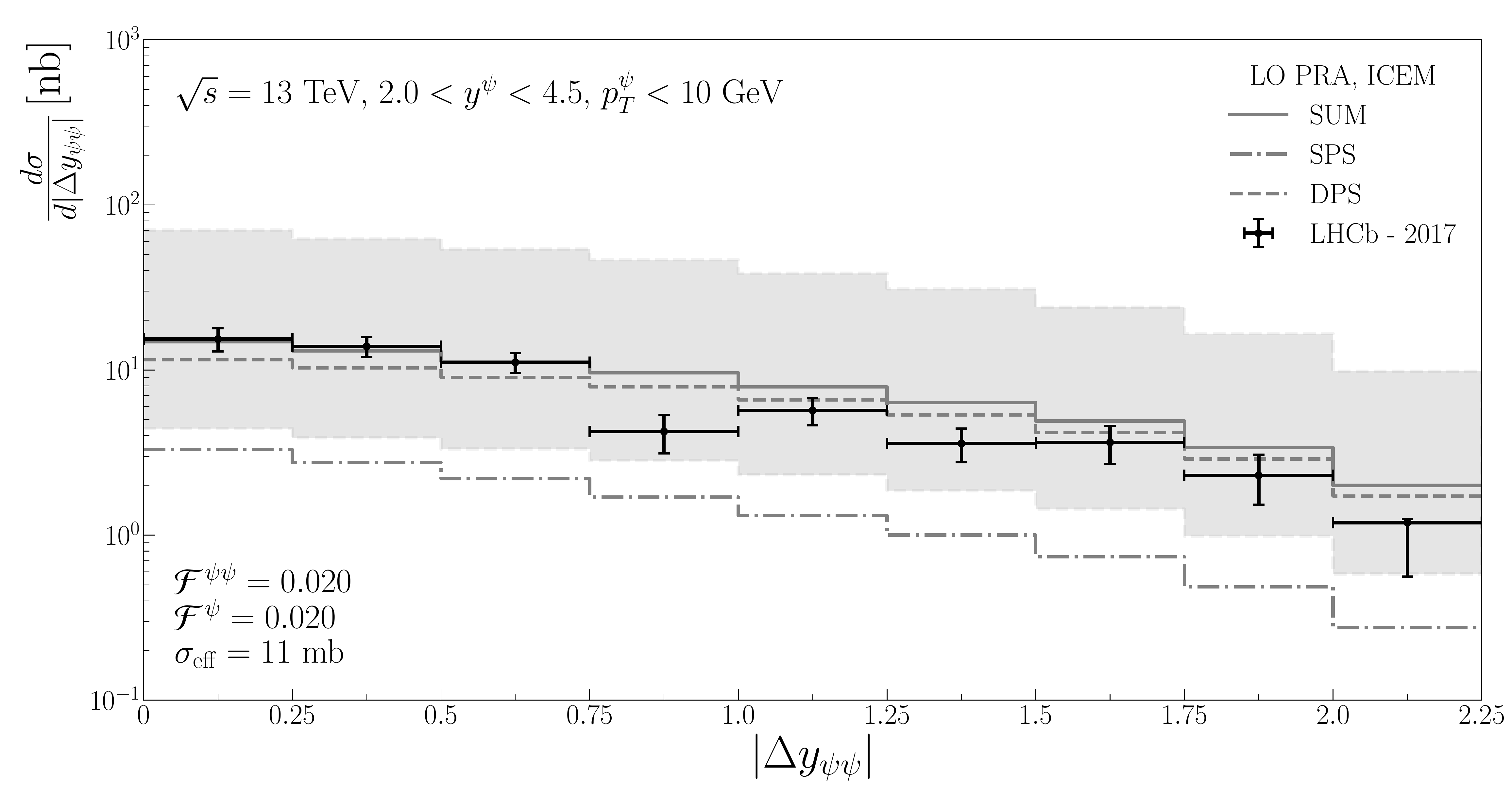}
\includegraphics[width=0.45\textwidth,angle=0]{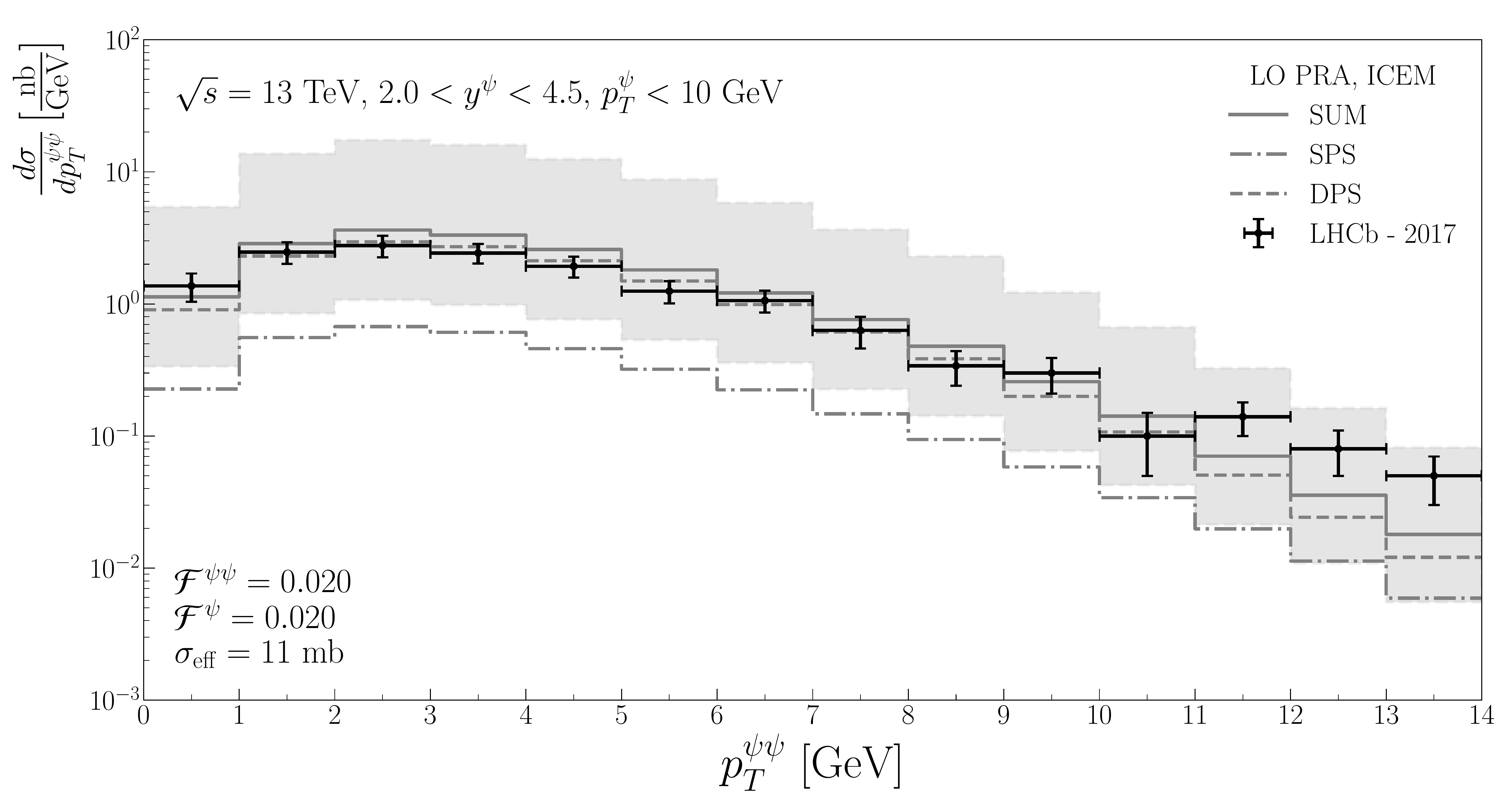}\includegraphics[width=0.45\textwidth,angle=0]{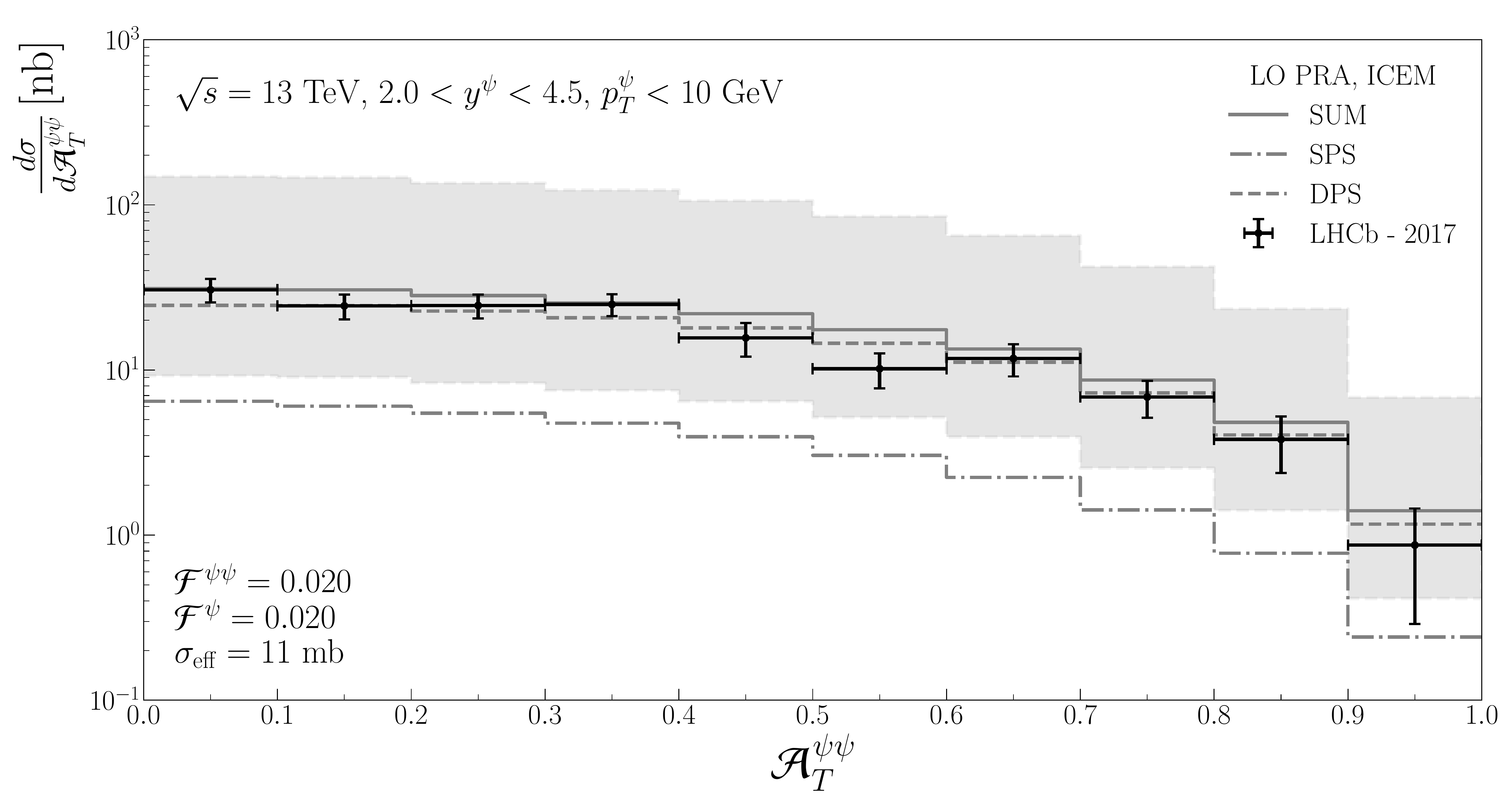}
\includegraphics[width=0.45\textwidth,angle=0]{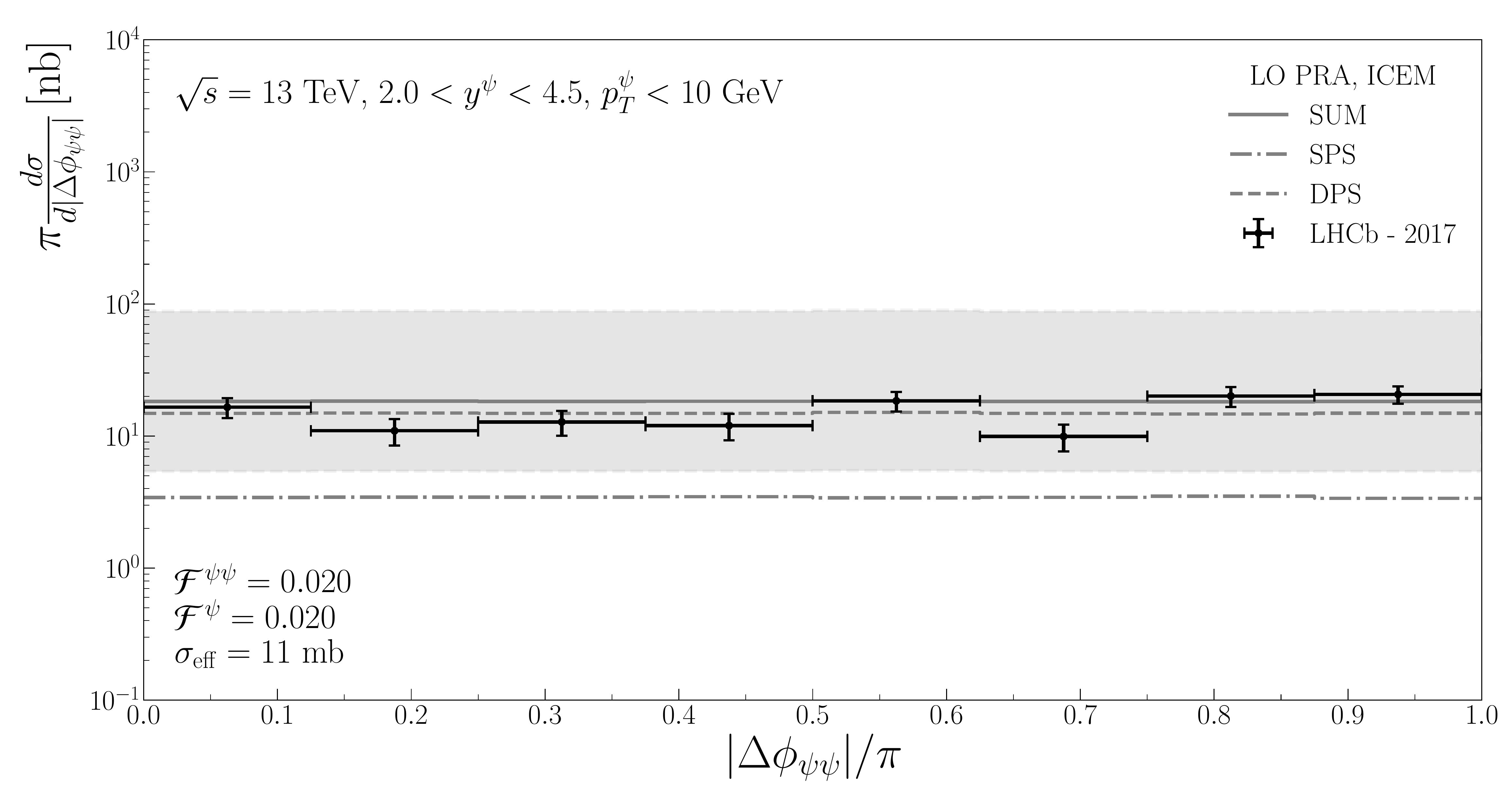}
    \end{center}
\caption{  Different spectra of pair $J/\psi$ production on
$m_{\psi\psi}$, $|\Delta y_{\psi\psi}|$, $p_{T}^{\psi\psi}$,
$A_T^{\psi\psi}$ and $|\Delta\phi^{\psi\psi}|$. The data are from
LHCb collaboration \cite{LHCb_2Psi}. \label{fig_LHCb2}}
\end{figure}

\begin{figure}[h]
\begin{center}
\vspace{0cm}
\includegraphics[width=0.45\textwidth,angle=0]{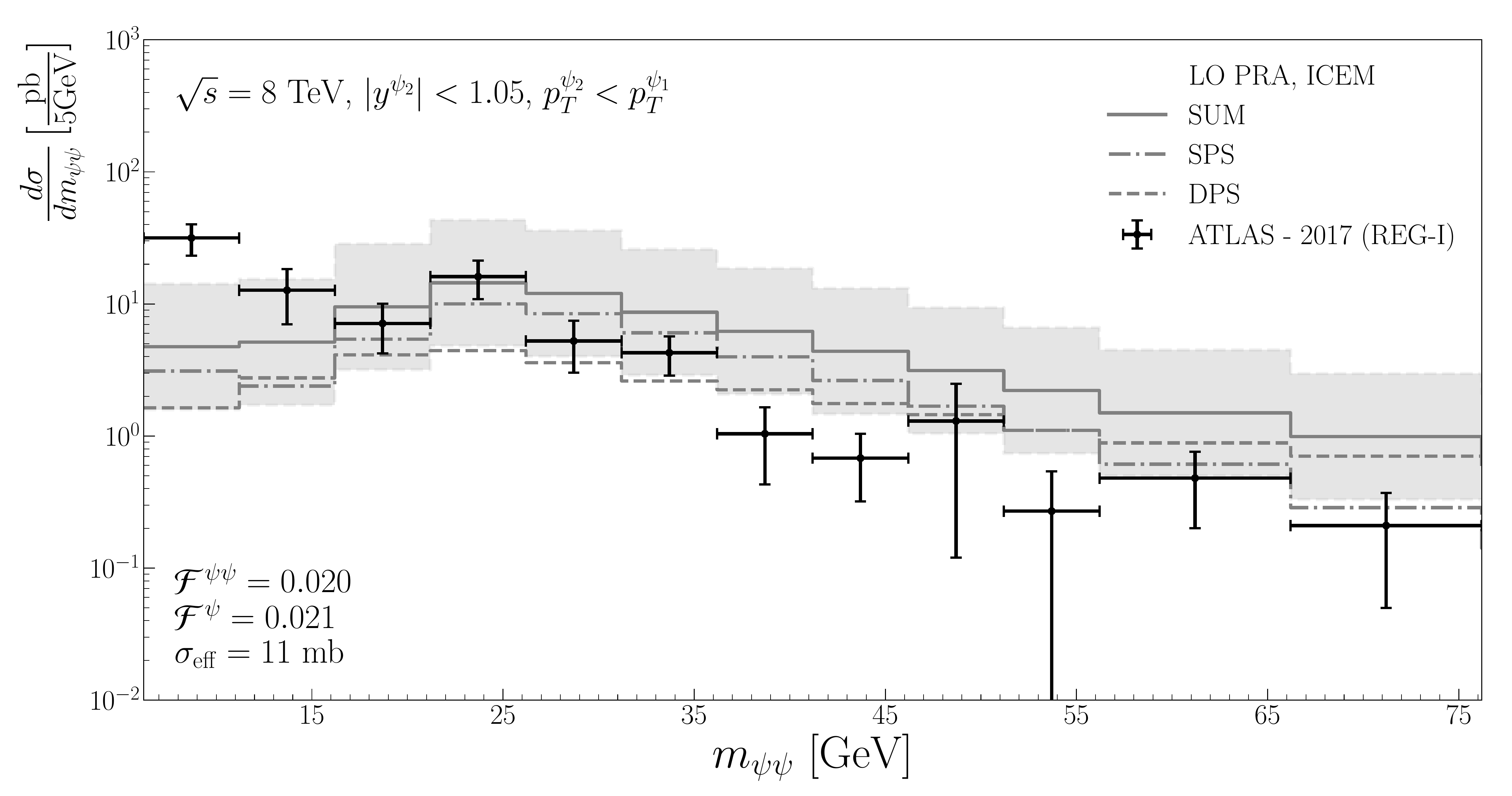}\includegraphics[width=0.45\textwidth,angle=0]{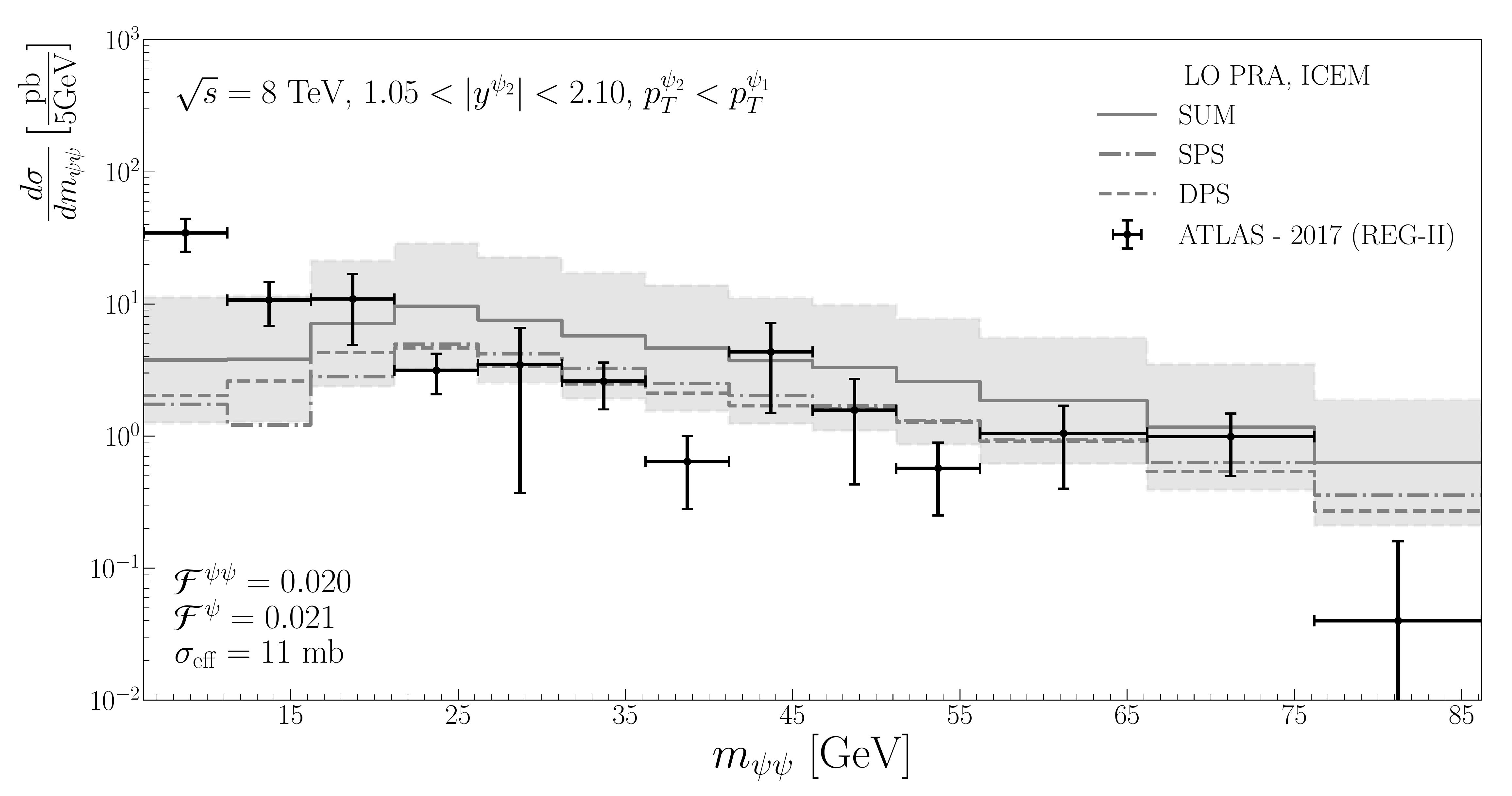}
\includegraphics[width=0.45\textwidth,angle=0]{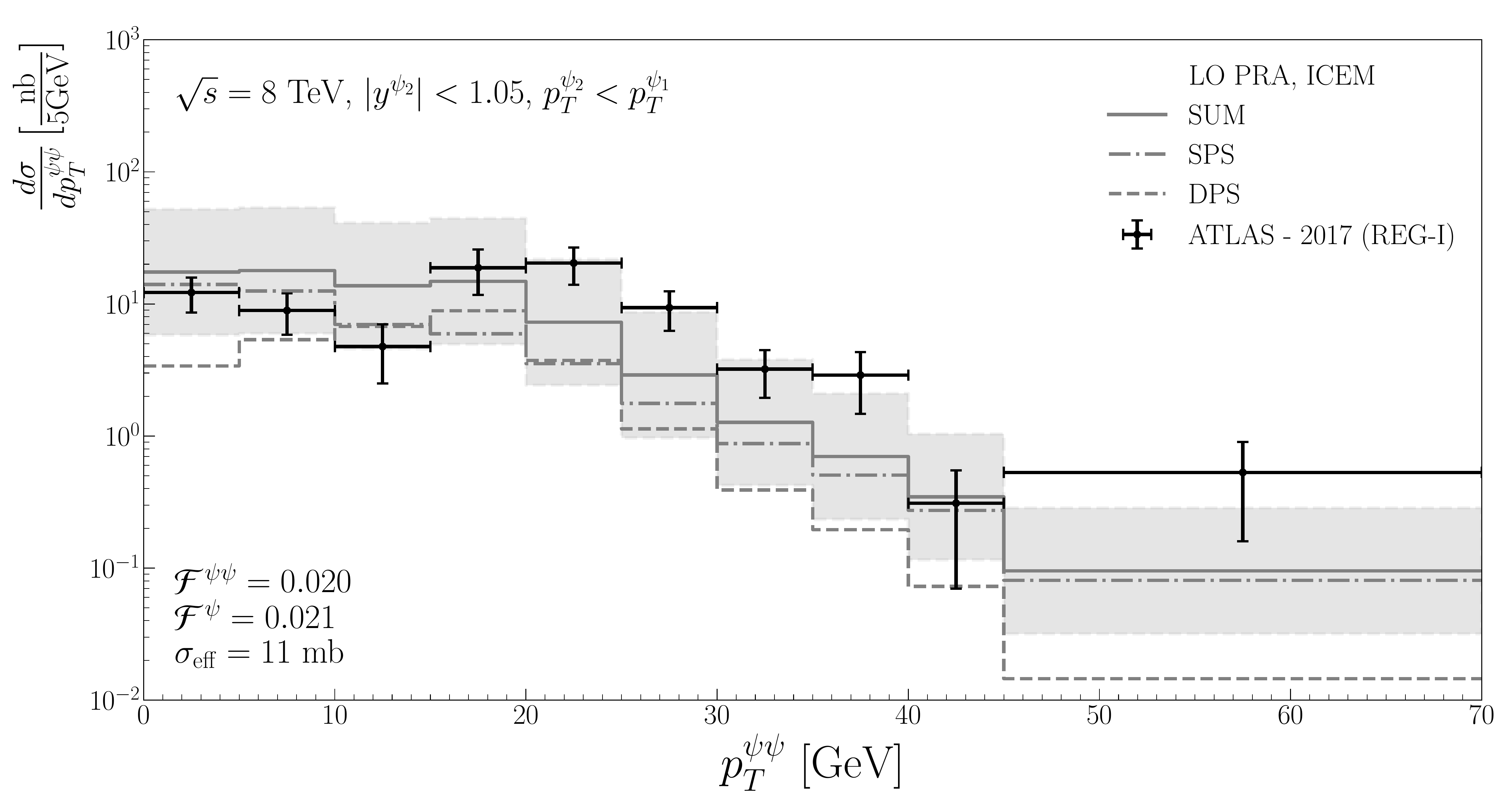}\includegraphics[width=0.45\textwidth,angle=0]{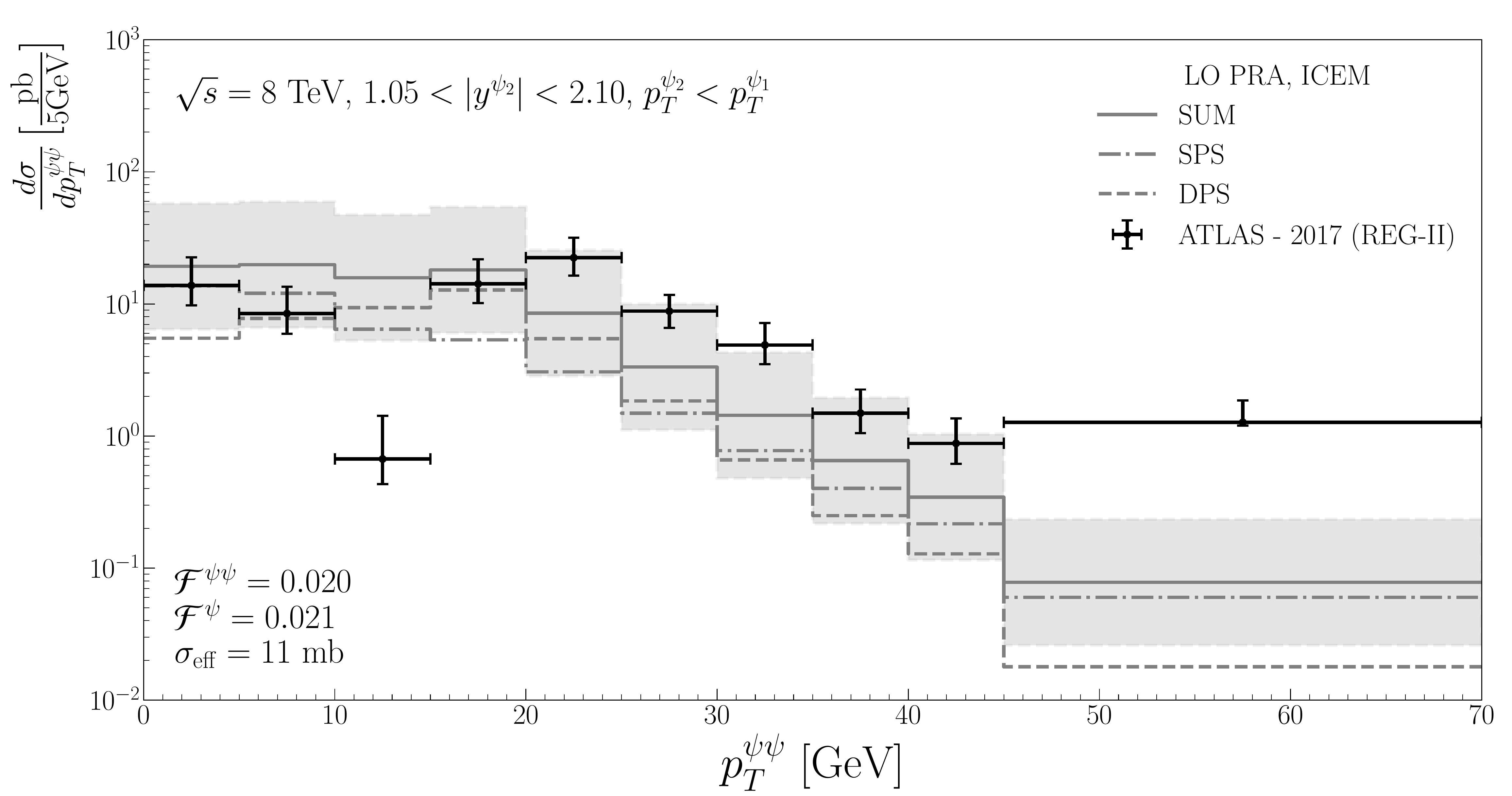}
\end{center}
\caption{Different spectra of pair $J/\psi$ production on
$m_{\psi\psi}$ and $p_{T}^{\psi\psi}$ for central and forward
rapidity regions. The data are from ATLAS collaboration
\cite{ATLAS_2Psi}}. \label{fig_ATLAS2}
\end{figure}

\begin{figure}[h]
\begin{center}
\vspace{0cm}
\includegraphics[width=0.45\textwidth,angle=0]{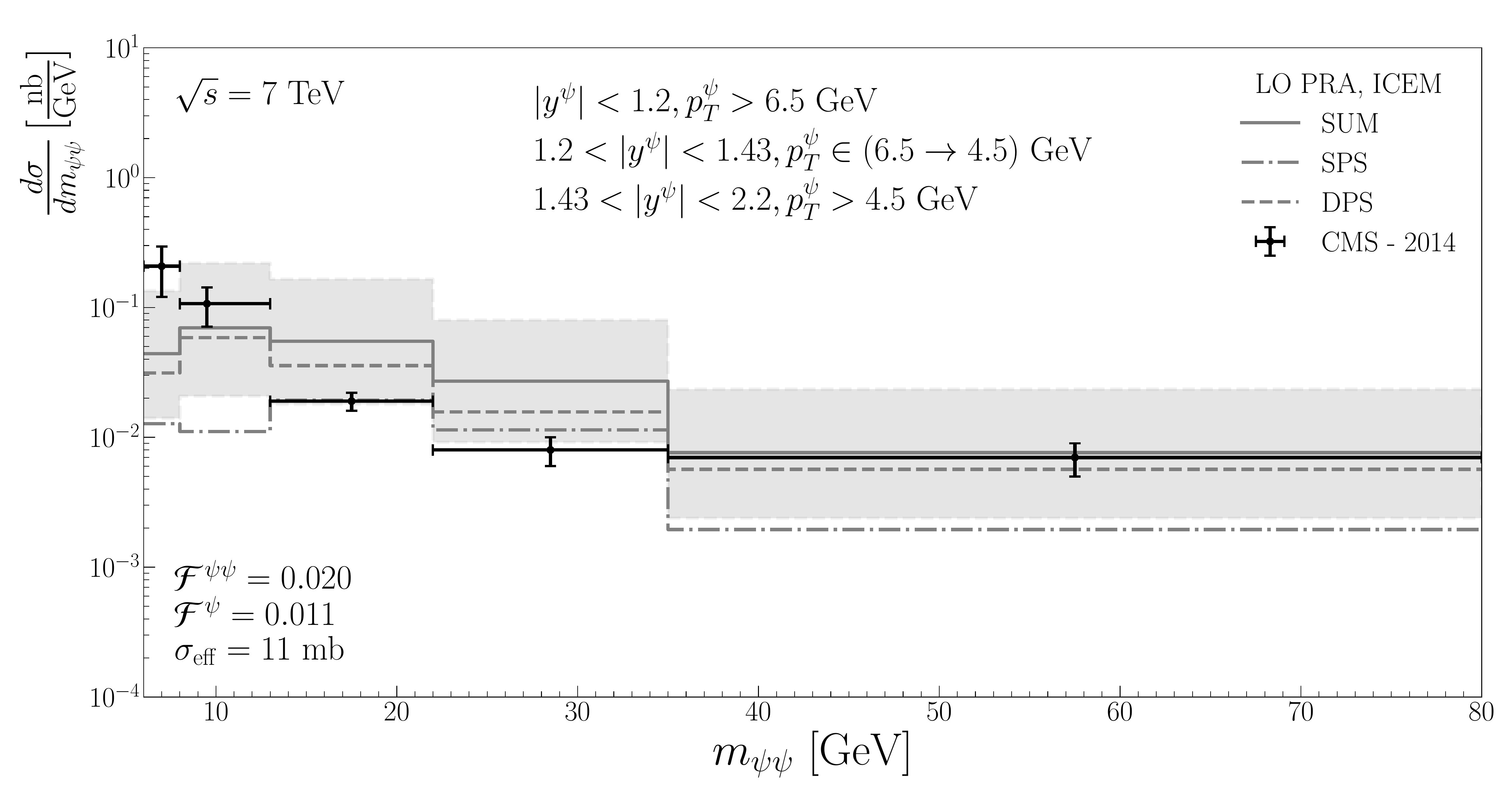}\includegraphics[width=0.45\textwidth,angle=0]{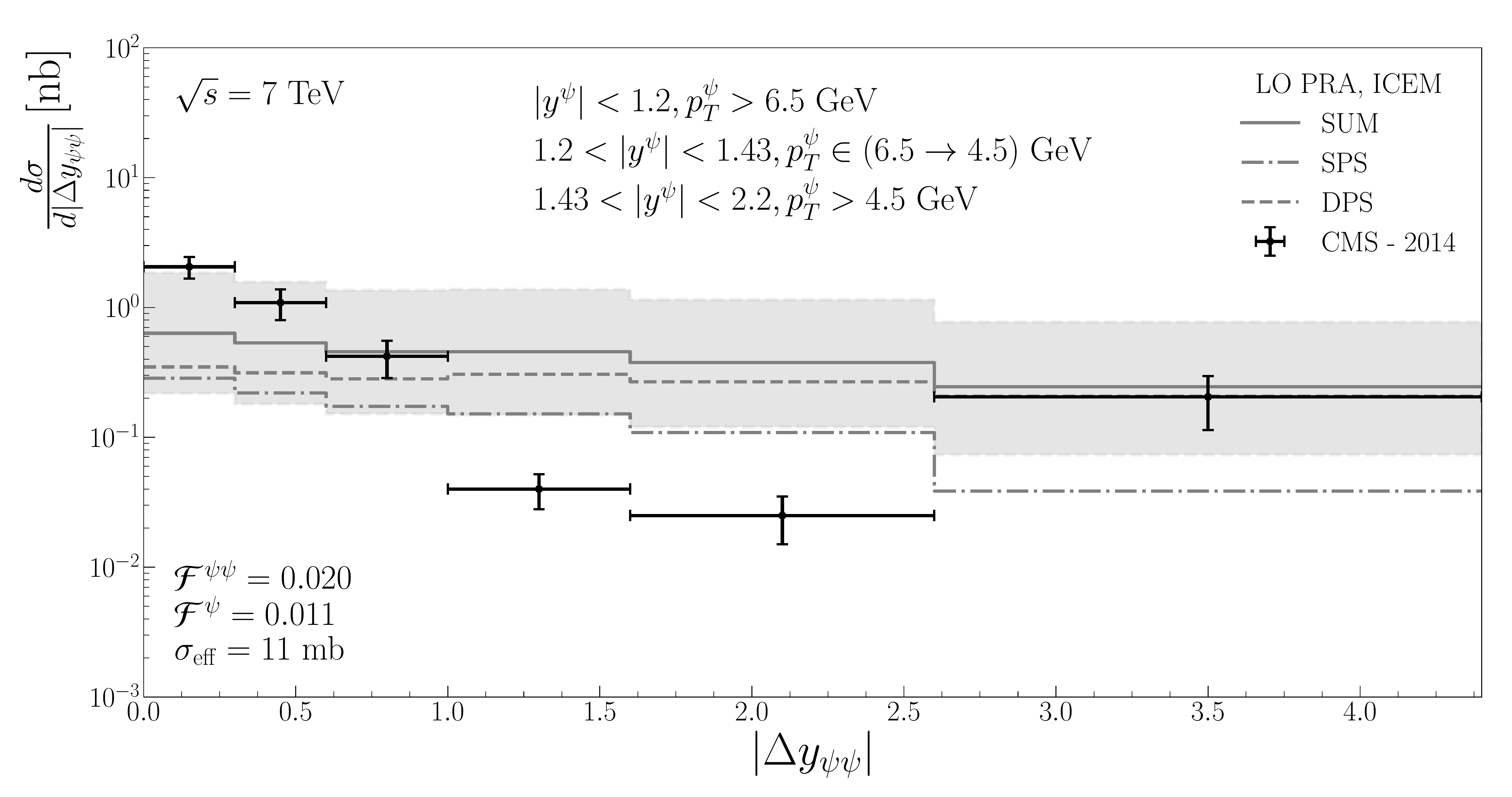}
\includegraphics[width=0.45\textwidth,angle=0]{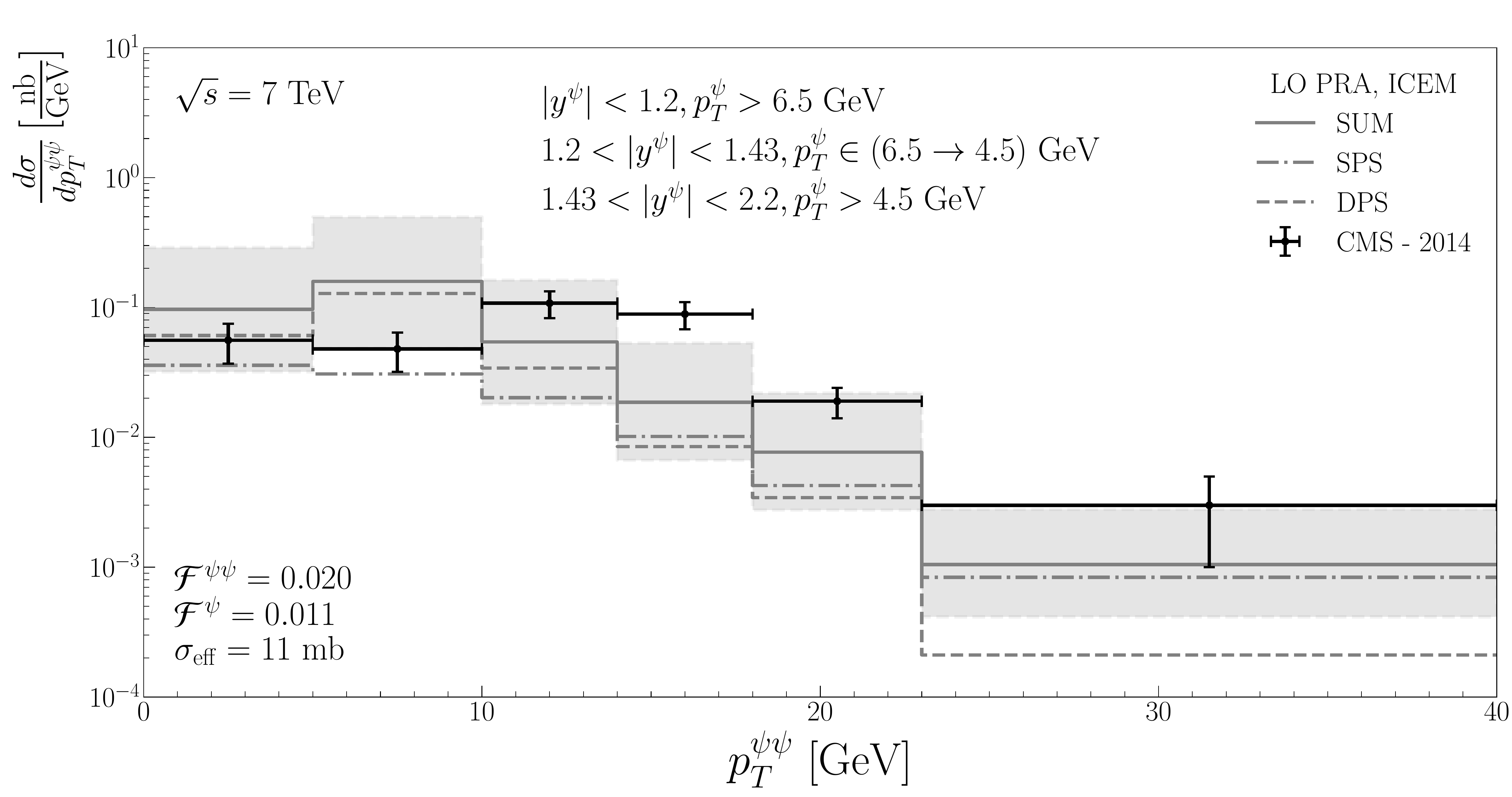}
\end{center}
\caption{Different spectra of pair $J/\psi$ production on
$m_{\psi\psi}$, $|\Delta\phi^{\psi\psi}|$ and $p_{T}^{\psi\psi}$.
The data are from CMS collaboration \cite{CMS_2Psi}}.
\label{fig_CMS2}
\end{figure}

\section{Conclusions}
\label{sec:conclusions}

We obtain a quite satisfactory description for the single prompt
$J/\psi$ $p_T-$spectra and cross sections in the ICEM using the PRA
at the wide range of the collision energy. The obtained values of
the hadronization parameter $\mathcal{F}^\psi$ depend on energy, and
such dependence can be approximated by the formula
$\mathcal{F}^\psi(\sqrt{s})=0.012+0.952 (\sqrt{s})^{-0.525}$. The
exact physical interpretation of such energy dependence needs
special analysis.

Both mechanisms, SPS and DPS, for the pair $J/\psi$ production have
been considered. We show the assumption
$\mathcal{F}^{\psi\psi}=\mathcal{F}^\psi \times \mathcal{F}^\psi$ is
not correct in the ICEM, and we find  $\mathcal{F}^{\psi\psi} \simeq
\mathcal{F}^\psi$ at the high energy.

The data for the pair $J/\psi$ production cross sections at the
energy range $7-13$ TeV can be fitted self-consistently with two
free parameters ${\mathcal{F}}^{\psi\psi}$ and $\sigma_{eff}$. We
have found  the best fit with ${\mathcal{F}}^{\psi\psi}\simeq 0.02$
and $\sigma_{eff} \simeq 11.0$ mb, when parameter
${\mathcal{F}}^{\psi}$ is fixed independently in the study of the
single $J/\psi$ production. We find the dominant role of the DPS
mechanism only in the case of forward pair $J/\psi$ production. At
the central region of $J/\psi$ rapidities, both mechanisms
contribute approximately equally.

\section*{Acknowledgments}
We are grateful to A. Van Hameren for advice on the program KaTie;
M. Nefedov, A. Karpishkov and A. Shipilova for helpful discussion.
The work was supported by the Ministry of Science and Higher
Education of the Russian Federation, project FSSS-2020-0014.

\clearpage

\bibliography{bibliography}

\end{document}